\documentclass[%
aps,
prx, 
amsmath,amssymb,
preprint,%
superscriptaddress,
altaffilsymbol
]{revtex4-2}

\usepackage{graphicx}
\usepackage{dcolumn}
\usepackage{bm}

\usepackage[utf8]{inputenc}
\usepackage[T1]{fontenc}
\usepackage{etoolbox}
\usepackage[dvipsnames]{xcolor}

\newcommand*{\Comb}[2]{{}^{#1}C_{#2}}%
\usepackage{algorithm}
\usepackage{algpseudocode}
\usepackage{chemformula}
\usepackage{amsmath}
\usepackage{ulem}
\usepackage[version=3]{mhchem} 
\usepackage{footmisc}

\newcolumntype{P}[1]{>{\centering\arraybackslash}p{#1}}

\makeatletter
\def\@email#1#2{%
 \endgroup
 \patchcmd{\titleblock@produce}
  {\frontmatter@RRAPformat}
  {\frontmatter@RRAPformat{\produce@RRAP{*#1\href{mailto:#2}{#2}}}\frontmatter@RRAPformat}
  {}{}
}%
\makeatother
\begin{document}


\title{Algorithmic detection of crystal structures from computer simulation data}
\author{Sumitava Kundu}
\altaffiliation[Current address: ]{Department of Chemical Engineering, University of Michigan, Ann Arbor, USA, MI 48109-2800}
\affiliation{School of Chemical Sciences, Indian Association for the Cultivation of Science, Kolkata, India, 700032}
\author{Kaustav Chakraborty}%
\affiliation{School of Chemical Sciences, Indian Association for the Cultivation of Science, Kolkata, India, 700032}
\author{Avisek Das}
\thanks{Corresponding author}
\email{mcsad@iacs.res.in}
\affiliation{School of Chemical Sciences, Indian Association for the Cultivation of Science, Kolkata, India, 700032}

\date{\today}

\begin{abstract}
Detection of crystal structures from particle positions of crystalline assemblies formed in computer simulations is an unsolved problem. The standard protocol, formulated in the reciprocal space, for structure determination from experimental diffraction data is not suitable for analysis of computer simulation data, after converting them to the Fourier space. This is primarily due to the extremely small sizes of the simulation systems compared to macroscopic crystals used in diffraction experiments. There is a long history of attempts to tackle this problem by analyzing the system in the real space by using ideas of local neighbors and broken symmetries of the crystalline state. But all these approaches fail for complex crystals with multiparticle basis. In this paper, we propose a heuristic solution to this problem by detecting all possible unit cells directly from particle coordinates obtained in a typical computer simulation. The method is based on well known facts about crystal structures, some of which are underutilized in the context of the current problem. These include, the symmetry of the coordination polyhedron and its empirical relationship with directions of lattice vectors for a simple Bravais lattice, and the fact that any complex crystal can be systematically decomposed into multiple Bravais lattices. By using these ideas, along with standard computational techniques like search, clustering and convex hull construction, we were able to handle complex basis and construct all crystallographically viable unit cells from the coordinates. The method is capable of handling statistical noise by employing certain cutoffs and deals with multicomponent systems in a transparent manner. We validated it on real Monte Carlo simulation data and variety of test systems, including crystals with tens of particles in the basis. Our heuristic algorithm, which requires minimal human intervention and computational resources, provides a solution to the long standing problem and would be beneficial to the wider communities of condensed matter physics and computational materials science. 
\end{abstract}
\maketitle

\section{\label{sec:Introduction}Introduction} 
Determination of microscopic structure of any substance is of utmost important both from fundamental and practical standpoints. It provides the true, unambiguous identity of the material and a framework for understanding its properties, eventually leading to targeted applications. For a single crystalline material, the structure is well defined and expressed in terms of the lattice parameters and the space group of the corresponding crystal structure. Structural elucidations of atomic and molecular materials belonging to this class are routinely carried out by x-ray crystallography \cite{Giacovazzo1992, Massa2004, Xie2013, Katrin2016, Deng2006, Dubach2020, Computational1994, Sheldrick2015, Dolomanov2009}. The diffraction data is collected in the reciprocal space and the physical unit cell of the crystal, meaning the lattice vectors and the effective coordinates of the atoms in the basis in the real space, are derived by an iterative refinement procedure, after taking care of the sign problem. These methods appeared to be useful in crystals made out of larger building blocks, for example nanoparticles. Only domain where a crystalline assembly can be analyzed based on particle coordinates pertains to the field of colloids. Because of larger size, the colloidal particles can be tracked directly and particle coordinates could be used as the basic input, rather than the diffraction data in the reciprocal space, for structural studies of colloidal crystals. Impressive advancement has been recently reported in this area \cite{Zang2024}. 

The problem of crystal structure detection arises in computational studies in a completely different setting. Unlike most real experiments, here particle coordinates are readily available and the task is to construct the unit cell from the real space particle coordinates. This problem may not have a unique solution, as it is well known that unit cells are not necessarily unique, when constructed from positions of the particles in the crystal. While it is tractable for simple crystals, for complex ones with many particles in the basis, the objective is quite challenging even for experts, and is extremely difficult for a normal user, which is the case for many simulations studies. Even if a solution that primarily depends on manual inspection could be found out, repeating that for large volumes of data is quite difficult. A better option is to look for algorithmic solutions. It is not possible to use the techniques employed in experimental solution of crystal structures in the reciprocal space, even though real space coordinates from simulations can readily be converted into structure factors in the reciprocal space. The reason behind this is the small system sizes used in the simulations. Reciprocal space data from such small systems are not adequate for procedures that rely on diffraction data from bulk crystals \cite{Engel2021}. Over the years, with increasing computational power, computer simulations became an extremely important tool in diverse areas of natural sciences where the outcomes of ``computer experiments'' are formation of crystalline aggregates. For example, in the computational studies of spontaneous self-assembly, where the primary objective is to understand how properties of building blocks control the final outcome of the assembly \cite{Damasceno2012, Agarwal2011, OlveraDeLaCruz2016, Li2016, Boles2016}. If those are crystals, identification of the structure is the first order of business, and one needs to do that efficiently on large data sets produced in typical studies carried out in modern computational materials science. The problem of detection of crystal structure from simulation data has become a central issue in further advancement of this field with important implications in several allied areas of physics, chemistry and materials science.

This problem has been recognized and attempts have been made to solve it in computational research for decades. The first serious attempt towards making decisions whether some particles in a dense liquid phase were forming crystalline order was put forth by the pioneering work of Steinhardt \textsl{et al.}\,\cite{Steinhardt1983}. They introduced the crucial notion of bond orientational order. These ideas have been further extended by several investigators into systematic methods for identification of crystal structures from computer simulation data \cite{John2022, Eslami2018, Logan2023, mukhtyar2018, mukhtyar2022}. Another approach that relied on local neighbors of particles, called common-neighbor analysis (CNA) \cite{Honeycutt1987, Faken1994}, has also been quite influential and other methods have been developed around CNA, for example, templating \cite{Duff2011} and graph based structural analysis protocols \cite{Reinhart2018, Li2023}. A different technique called polyhedral template matching (PTM) was introduced which has been successful in treating quite a few crystal classes \cite{Larsen2016}. Recent machine learning based algorithms appeared to be extremely successful in determining the structural classes of crystallites formed in simulations, but they typically required a large amount of local fingerprint data for training purposes. In these studies, various descriptors were proposed based on symmetry \cite{Ziletti2018}, smoothened versions of particle coordinates, for example, in smooth-overlap-of-atomic-positions (SOAP) \cite{Leitherer2021} and topology of the crystallographic environment \cite{Boattini2019}. Besides these, multiple developments successfully distinguished complex structures \cite{Spellings2018, Boattini2018}, but failed to provide specific crystallographic information without proper references. Many useful techniques have been developed to detect structures of molecular systems based on the chemical spaces \cite{Boles2016, Bhattacharyya2018} or by tuning the crystalline behavior in the interaction spaces \cite{Rossi2011}. Recently, a theoretical approach was reported that analyzed the point group of the local neighbor directly based on Weigner \textsl{D}-matrices that shed light on the paths towards point group detection of crystal structures \cite{Engel2021}. The rotational symmetry of the neighbor was detected, but it was unable to reveal the complete crystallographic symmetry and the space group. Considering the limitations of the existing techniques, it is fair to state that identification of the crystal structure in terms of complete information about the unit cell and the space group still requires a lot more attention, and should be regarded as an open problem.

In this paper, we report a direct approach to detect all possible unit cells and the corresponding space group from particle coordinates of crystallites formed in computer simulations. The algorithm was based on several known facts about analysis of crystalline assemblies. These include local environments, Bond Orientational Order (BOO) diagram and coordination polyhedron. It also used several other aspects of crystal structures that had not being used so far in the current context. In this regard, a key insight which played a vital role was the realization that any complex structure could be systematically processed into multiple crystals consisting of subsets of the original particles. Each of these subsystems formed a simple Bravais lattice which had specific relationship with the original system. We made extensive use of standard computational techniques, e.g.\,clustering, search and convex analysis, at appropriate steps of the algorithm. The heuristic method is presented in two parts. First we illustrate how the problem was solved for simple Bravais lattices, and then we provide the general solution for any kind of crystal with arbitrarily complex basis. The final output of our algorithm was all possible crystallographically viable unit cells, in other words, the lattice parameters and the positions of effective number of particles. The space group was deduced from this information by utilizing the widely used \textit{Spglib} package \cite{spglibv1}. The method was semi-automatic in nature and required human intervention only in terms of specification of certain cutoffs and visual inspections which could be carried out in a straightforward manner. It was able to deal with complex crystals with large bases and was quite robust in handling statistical noise present in simulation output.  

\section{Unit cells and anisotropic environments of particles in a crystal }\label{basic_concepts}
Here, we briefly describe some well-known facts about crystal structures and highlight few concepts most relevant for the ensuing discussion of our algorithm. In three dimensions, crystalline state is characterized by 7 crystal classes, 14 Bravais lattices and 230 crystallographic point groups. A full specification of a crystal involves identification of one of these space groups, which is consistent with the basis and the lattice parameters, in other words, the conventional unit cell of the crystal. In this algorithm, we focus on the detection of the conventional unit cell from the particle identity and coordinate data obtained from a typical computer simulation. However, it is well-known that, in such a setting, the choice of unit cell is not unique. We handled this issue by enlisting all possible choices of basis vectors and unit cell parameters that seemed to be compatible with the entire data as processed by our algorithm. While acknowledging this unavoidable fundamental limitation, we used all the obvious crystallographic restrictions that could be used to make the proper set of choices. In other words, the non-uniqueness problem was handled in such a way, the overall algorithm did not violate other restrictions imposed by the general properties of three dimensional crystals. The implications of these comments in the context of our heuristic algorithm will be apparent in the remainder of the paper.

\begin{figure}[!h]
	\centering 
	\includegraphics[scale=0.4]{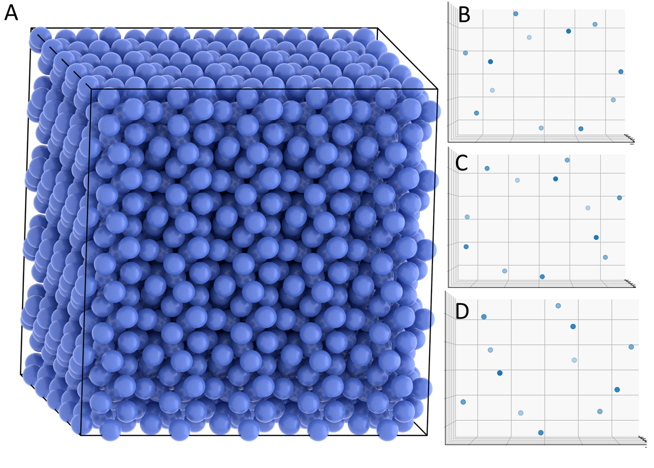}
	\caption{\textbf{Different kinds of positional environment in $\beta$-Mn crystal are shown.} An ideal structure of $\beta$-Mn crystal is shown in \textbf{(A)}. The distributions of neighbors within the distance $r_c$ $\sim$ 2.6 are shown for three arbitrarily chosen particles showing completely different kinds of distributions for the particles indicating the system as a non-Bravais lattice as shown in \textbf{(B)}, \textbf{(C)}, \textbf{(D)}.}
	\label{fig:env}
\end{figure}

\subsection{Local environments of particles in a crystal} \label{sec:local_env}
Both translational and rotational symmetries are broken in the crystalline state. As a result, the local neighborhood of each particle in a crystal is anisotropically populated by other particles. This is in contrast to the situation in  liquid state, where, over a sufficient duration of time, a particle is surrounded by others irrespective of the angular direction. Anisotropic ``local environments'' of particles can be formally defined, within a length scale characterizing the definition of local neighborhood, in the following way. The neighbors of a center particle with position $\vec{r}_i$ are defined as the set of all particles within a distance $r_c$. The number of such neighboring particles is denoted by the ``coordination number'', $N_{CN}$, which is dependent on the value of cutoff. The pairwise vectors, defined as $\vec{r}_{ij} = \vec{r}_j - \vec{r}_i$, with $i=1,2,\ldots,N$ and $j=1,2,\ldots,N_{CN}$, captures the local translational and rotational broken symmetries around the particle $i$. This idea is well-known and commonly referred to as the ``bond orientational order'', pioneered by Steinhardt \textsl{et al} \cite{Steinhardt1983}. The set of bond vectors, $\mathbb{G}_{i,r_c} = \{\vec{r}_{ij}\}$, can be used as a local descriptor of the crystal structure.

The set of bond vectors can be compared to detect whether the local environments of two particles, with indices $i$ and $j$, are same or not.  It should be noted that the magnitude of bond vectors are not unit and include the neighbor distances as depicted by the crystal structure. Besides that, the arrangements of bond vectors corresponding to the reference particle might have the site symmetry of a Wyckoff position, but the local environment of a particle directly refers to the coordination polyhedron as discussed in the literature \cite{Larsen2016}. It is important to remember, only similar Wyckoff positions are not considered here, instead of that, the entire local environments of particles are compared for the detection of different kinds of local symmetry breaking. This comparison only makes sense at a single value of cutoff $r_c$, and orders of elements in the sets $\mathbb{G}_{1,r_c}$ and $\mathbb{G}_{2,r_c}$ are ignored for this purpose. Directly plotting the sets in three dimensional Cartesian space provides a visual way to understand the difference in local environments between two particles in a crystal. An important fact about local environments is that for all fourteen Bravais lattices, with one particle at each lattice site, it is always possible to find out the particles with same local environments at any cutoff distances. For a simple system with the identical particle types, the cutoff can easily be estimated from the first peak of the radial distribution function (RDF). Complex crystals have multiple local environments i.e., different sets of bond vectors $\mathbb{G}_{i,r_c}$, for different values of $i$ at any particular length scale defining the local neighborhoods \cite{Kittel2005, Bernstein2009}. A structure with one particle per lattice site where multiple local environments exist, will be referred to as a non-Bravais lattice. For example, in the structure of $\beta$-Mn, three different environments are shown in Figs.\,\ref{fig:env}B, C, D within the cutoff distance $\sim$ 2.6. This local environments and their algorithmic detection from the particle coordinate data are key ideas of our overall heuristic crystal structure detection method as discussed in the subsequent sections.

\begin{figure*}[!h]
	\centering 
	\includegraphics[scale=0.3]{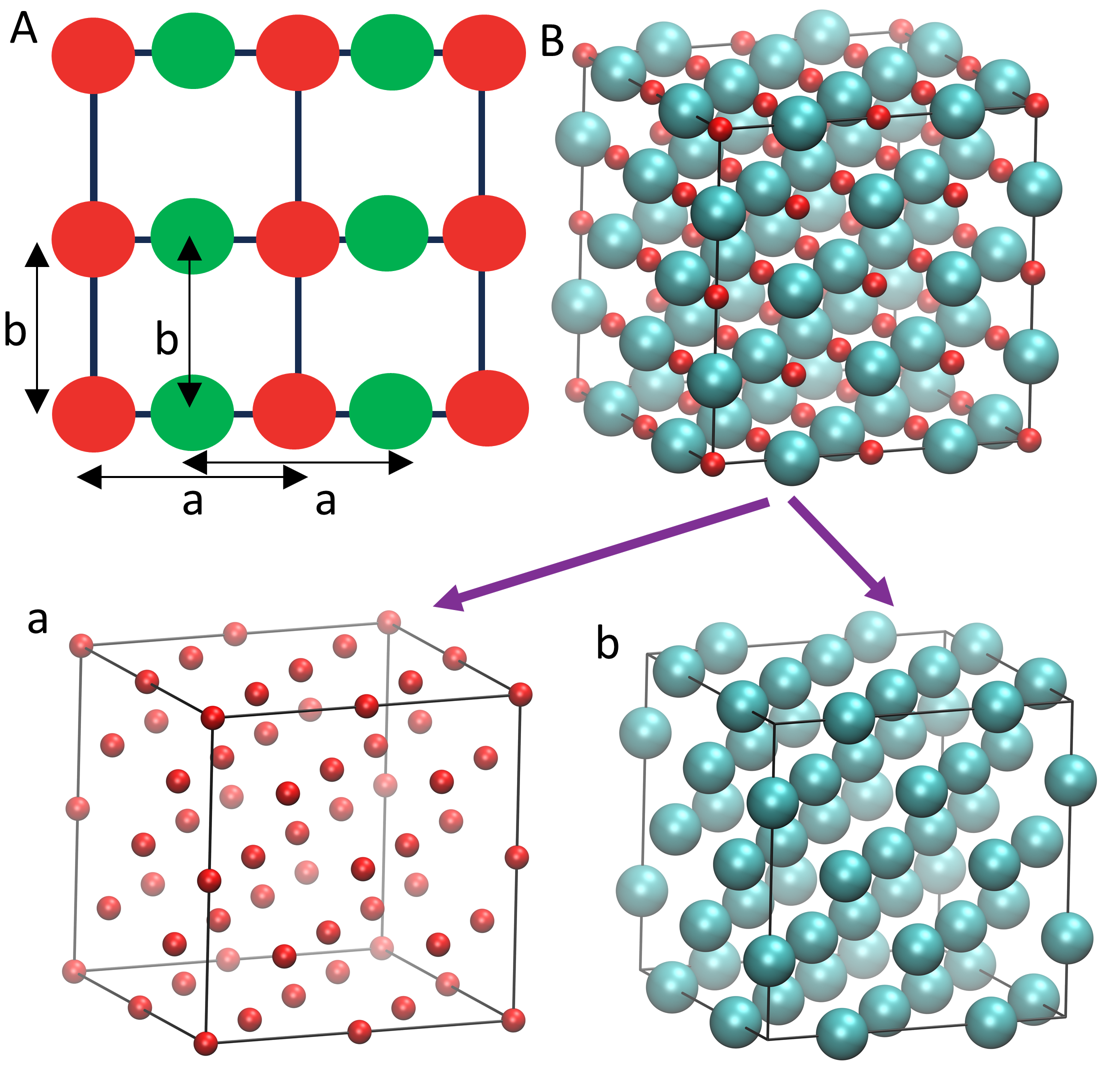}
	\caption{\textbf{Separation of particles based on identities is shown.}  A 2D cartoon diagram is presented in panel \textbf{(A)} to show the association of unit cell with either ``red'' particles or ``green'' particles leading to the same lattice parameters and basis vectors. Considering this fact, \ch{NaCl} crystal structure (shown in panel \textbf{(B)}) is categorized into two different systems depending on the atom type with panel \textbf{(a)} and \textbf{(b)} displaying the coordinates of \ch{Na} and \ch{Cl} atoms respectively.}
	\label{fig:nacl}
\end{figure*}

For a Bravais lattice, similarity of local environments based on the bond vectors only holds true if all particles in the system have the same identity. In a system with multiple types of particles, the definition must incorporate the species labels along with the particle coordinates. For example, in the rock salt (\ch{NaCl}) structure (see Fig.\,\ref{fig:nacl}), \ch{Na} and \ch{Cl} atoms are positioned in such way, the atom types appear to be important for obtaining the same local environment. This important point is discussed in details in the later sections.

\section{Detection of unit cell for a simple Bravais lattice from particle coordinates}\label{sec:brav_alg}
Before we present the full algorithm for crystal structure detection via determination of all possible valid unit cells for complex crystals, we describe the method for a simple system. The simplest, text book example of a crystal is a single component Bravais lattice with single particle basis, meaning, each lattice site is occupied by one particle. This fact could be exploited by transforming the crystal into a well defined and unique geometric object which captured the broken symmetry of the crystal. This geometric object could easily be constructed by clustering, and in the next step, a sequence of searches, that respected the relevant crystallographic restrictions, could be performed to determine the crystal class, and eventually lattice parameters of all possible unit cells. This approach can be illustrated in the simplest crystal systems. The simple example also demonstrates how the non-uniqueness of the unit cell problem is dealt with within our geometric construction and search based protocol. This approach, although looks quite different from all the existing attempts, is closely related to many of them. The whole approach works because of the finiteness of the possible types of crystals in terms of seven crystal classes and fourteen Bravais lattices. Even though there could be infinite number of crystals even with the simple basis, these finite categorization made it possible to uniquely determine the essential features of the broken symmetries in terms of the aforementioned geometric construction. 

\begin{figure}[!h]
	\centering 
	\includegraphics[scale=0.2]{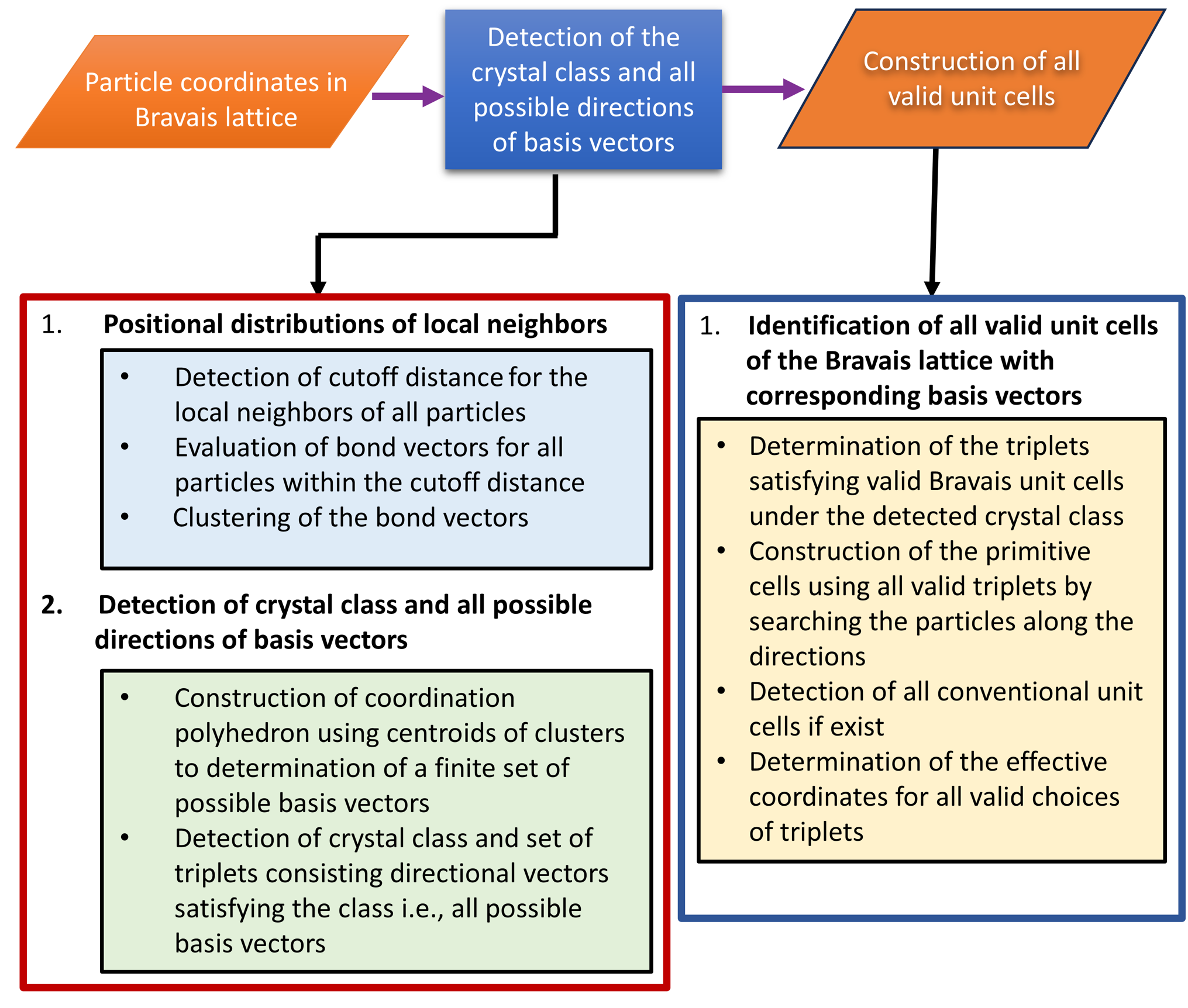}
	\caption{A schematic diagram illustrating the important steps of protocol is shown.}
	\label{fig:schematic_diag}
\end{figure}

Apart from the simplest set-up to present our method, unit cell detection of a simple Bravis lattice is an important step in the final algorithm for general crystalline systems. This fact will become clearer in the next section. In the rest of this section we describe the steps to identify the all possible unit cells of a crystal which is a Bravis lattice with one particle basis. For a summary of the entire algorithm for this part, refer to the flow chart in Fig.\,\ref{fig:schematic_diag}.  For the demonstration of the algorithmic steps, a simple cubic (SC) structure with lattice parameters $a$=6.315, $b$=6.315, $c$=6.315, $\alpha$=90$^{\circ}$, $\beta$=90$^{\circ}$, $\gamma$=90$^{\circ}$ has been used to describe each step of the algorithm towards the identification of the unit cell.  For validation purpose, the lattice parameters and the corresponding basis vectors of all possible unit cells obtained from our procedure were compared with the input values.

\begin{figure}[!h]
	\centering 
	\includegraphics[scale=1.0]{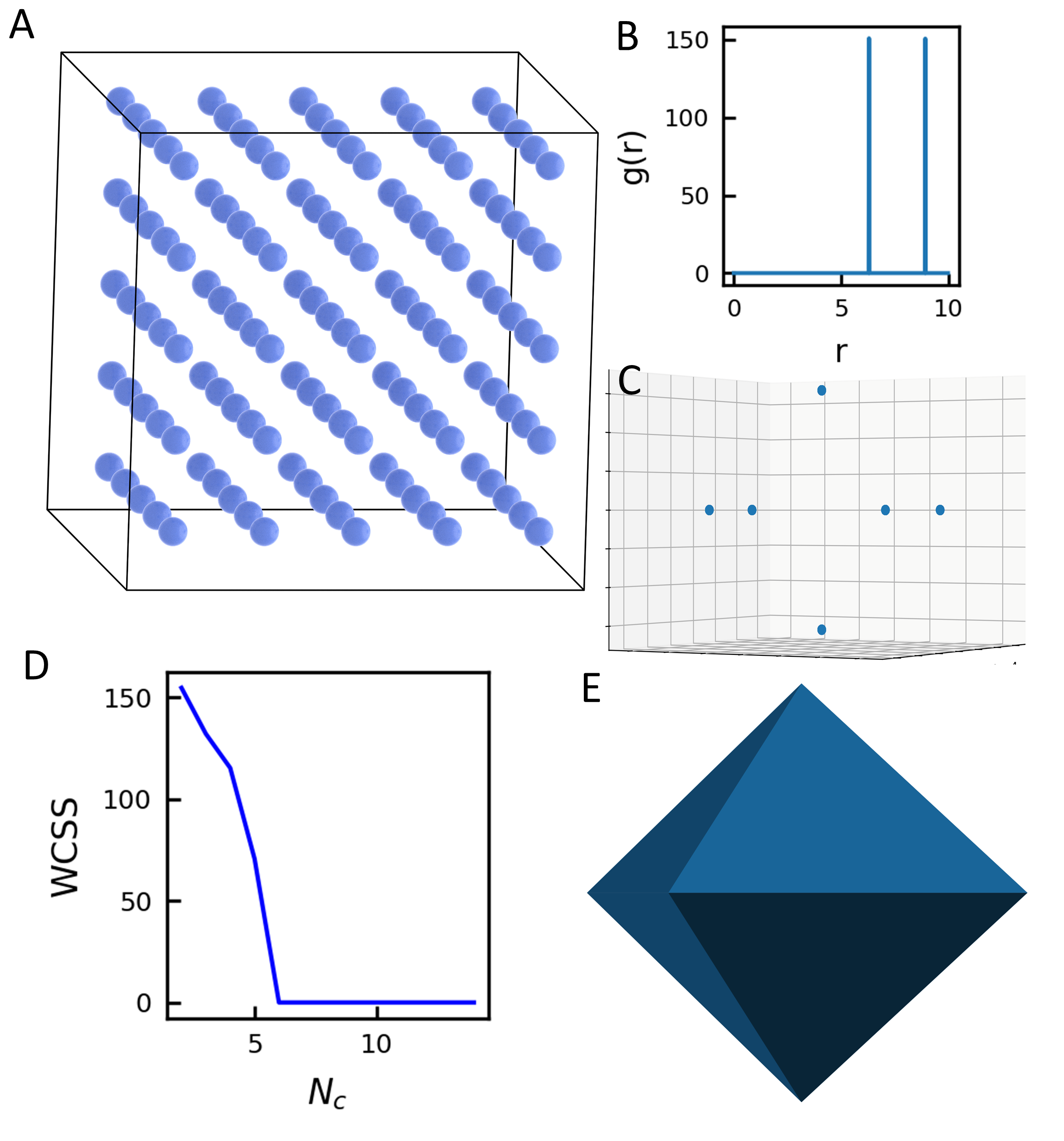}
	\caption{\textbf{Local positional environments and detection of clusters are shown for a simple cubic (SC) crystal structure.} An ideal system of SC crystal is shown in panel \textbf{(A)} with the corresponding RDF shown in panel \textbf{(B)}. Detecting the nearest neighbor distance $r_c$ $\sim$ 7.0, the BOO is displayed in panel \textbf{(C)} indicating the existence of six clusters of particles as confirmed by the \textit{K-Means} algorithm displayed in panel \textbf{(D)}. Panel \textbf{(E)} shows the constructed convex polyhedron using the centroid of the clusters.}
	\label{fig:convert}
\end{figure}

\subsection{Positional distributions of the local neighbors}\label{sec:alg_posi_dist}
The positional distributions of local neighbors were determined by the method described in the Section \ref{sec:local_env} after choosing the cutoff distance $r_c$. This is a straightforward process for a Bravais system as all particles have the same local environment i.e.\,all $\mathbb{G}_{i,r_c}$ sets were identical.  The bond vectors $\vec{r}_{ij}$s were superimposed around the center of the coordinate system, which gave a pictorial representation of the positional distribution of local neighbors. This is commonly referred to as the ``Bond Orientational Order'' (BOO) diagram or the bond order diagram. The detail has been described in the section 5.3 of Supplemental Material. The use of BOO diagrams for classification and detection of simple crystal structures has been quite common in many areas of condensed matter physics and material science, especially in computational self-assembly into crystalline states \cite{Steinhardt1983, Damasceno2012, Agarwal2011, Gantapara2013}. 

For an ideal crystal, the distribution of particles led to clusters with zero positional deviation, but in real simulation data with noise, multiple clusters within the statistical noise were observed. For automatic determination of the number clouds of points in a real simulation data we employed standard K-means clustering method. The number of clusters ($N_{c}$) was calculated using the ``elbow analysis'' of the two-dimensional plot of the sum of square distance between particles in a cluster and the cluster centroid (WCSS stands for ``Within-Cluster Sum of Square'') versus the k-value (\textit{x-axis}) as described in section 5.4 of Supplemental Material. The data for the test system is shown in Fig.\,\ref{fig:convert}D. The elbow was determined as the integer number in the \textit{x-axis} after which the value of WCSS decreased almost monotonically; the value of $N_{c}$ was 6. For a Bravais lattice, $N_{c}$ appeared to be equal with the coordination number $N_{CN}$ as there exists only one kind of environment for all particles in the system. The cutoff distance $r_c$ was set at $\sim 7$, which corresponded to the first peak of the RDF.

The implementation of this important but simple step in practice, however, could be problematic for the following reason. Making a proper choice of $r_c$ is less straightforward than one would imagine. The RDF, as used in the example system, could not be helpful and an improper choice could result in BOO which existed on a two dimensional manifold, therefore, being completely useless in identification of the local broken symmetries of the three dimensional crystals under investigation. This situation is most likely to arise for crystals with highly unequal lattice parameters, i.e.\, if the $c$ value is three times of the values of $a, b$ parameters. This situation could not be avoided in an automatic way and manual intervention was needed. This was not a serious issue as incorrect choice resulted in a lack of three dimensional structure of the neighbors, after all pairwise vectors were superimposed in a coordinate center at (0, 0, 0). In that case, the value of $r_c$ was required to be extended beyond the first minimum of RDF, which required manual adjustments towards progressively higher values. An unnecessarily large value of $r_c$ must be avoided, which could result in slower convergence of the subsequent steps.

\subsection{Determination of the crystal class and possible directions of basis vectors} \label{sec:alg_crys_class}
The next order of business was to determine the crystal class and directions of basis vectors for the Bravais lattice. This was done by analyzing the BOO with carefully designed search protocol. The two main sub-steps of this part of the algorithm are described below.

\subsubsection{Construction of a convex polyhedron from the positions of the local neighbors and detection of crystal class}\label{sec:alg_poly}
\begin{figure*}[!h]
	\centering 
	\includegraphics[scale=0.2]{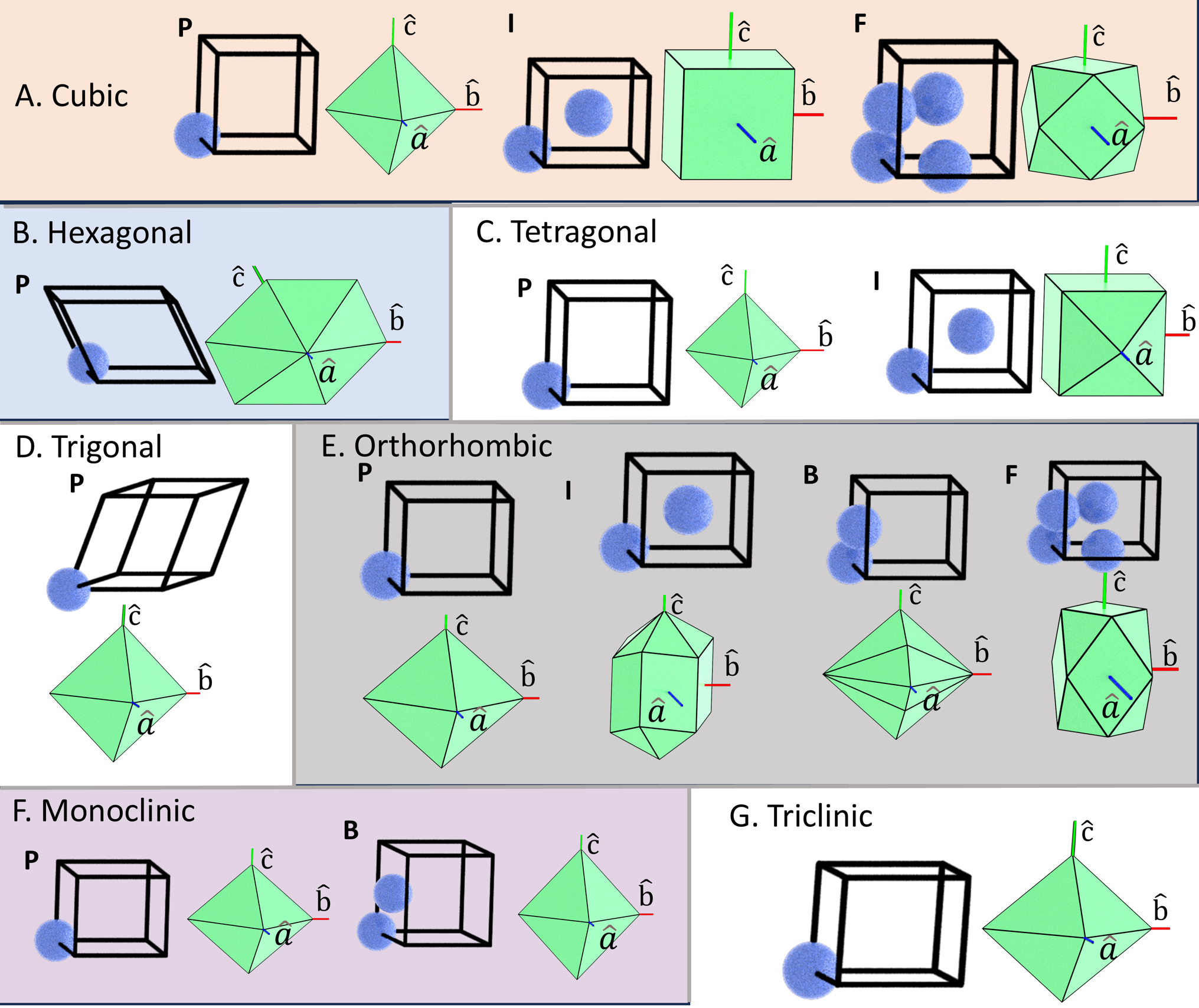}
	\caption{\textbf{Directions of the basis vectors are shown for all fourteen Bravais lattices .} Synthetically prepared unit cell of each Bravais lattice corresponding to \textbf{(A)} Cubic, \textbf{(B)} Hexagonal and \textbf{(C)} Tetragonal, \textbf{(D)} Trigonal, \textbf{(E)} Orthorhombic and \textbf{(F)} Monoclinic and \textbf{(G)} Triclinic crystal systems are displayed along with the respective coordination polyhedron. The directions of basis vectors, $\hat{a}$, $\hat{b}$, $\hat{c}$ are indicated for all cases. The letters \textbf{P}, \textbf{I}, \textbf{B}, \textbf{F} designate the abbreviations of primitive, body-centered, any base-centered and all face-centered systems respectively for each crystal class. For crystal classes except the cubic one, the geometry of coordination polyhedron could change but the directions of basis vectors appeared to intersect either the vertices or face mid-points or edge mid-points. Different colors reflect better visual representation of each section.}
	\label{fig:bravais_1}
\end{figure*}

Connecting the centroids of clusters in the BOO produced a convex polyhedron, which was not necessarily regular and referred to as coordination polyehdron in the literature \cite{Larsen2016}. This was essential geometric construction of the subsequent search problems which would ultimately lead to the unit cell. Because of the uniqueness of local environments in a simple single component Bravais crystal, the polyhedron was also unique.  It was important to appreciate that both the inter-particle distances and the rotational symmetry were incorporated in the BOO diagram. This information was enough to detect the broken symmetry of crystal, which will be more transparent as the discussion progresses. The shapes of the coordination polyhedra for all fourteen Bravais lattices are shown in the Fig.\,\ref{fig:bravais_1}. The geometry of the polyhedron could change depending on the values of the lattice parameters corresponding to specific classes but the point group symmetry would not. The polyhedra encompassed the essential features of the local symmetry breaking of the crystals under question. Hence, measuring the point group symmetry of the convex polyhedron would inform on the corresponding crystal class of the Bravais lattice. Angular tolerances were imposed to deal with the noise in real simulated crystals. In ideal scenario, the point group symmetry would be equal to any of the seven crystalline point group $O_h$, $D_{6h}$, $D_{4h}$, $D_{3d}$, $D_{2h}$, $C_{2h}$ and $C_{i}$. The detection of crystal class enabled the route to proceed further towards the detection of the unit cell of the Bravais system.

Analysis of certain geometrical features of this polyhedra would provide an algorithmic path, consisting of searches, to the identification of the unit cell. A similar approach was reported earlier in the context of crystal structure detection by matching the templates of the polyhedra with the arrangements of local neighbors \cite{Larsen2016}. The previous approach lacked the general application and was limited in determining a few simple crystal structures. All these limitations were taken care of and carefully circumvented by following an intuitive approach which appeared to be valid for a Bravais lattice under any crystal class. In a real situations with statistical noise, the construction of the polyhedra could be problematic, but the convexity property must be satisfied, otherwise, the algorithm could not proceed. These technical issues, which did not have direct bearing on the essential conceptual ideas underlying the method, will be discussed in the Results section with real data. 

\subsubsection{Determination of possible triplets of basis vector directions}\label{sec:alg_triplets}		

Careful inspection of the polyhedron constructed from the local neighbors distribution of any of the fourteen Bravais lattices belonging to seven crystal classes revealed an interesting fact. The direction of original basis vectors of the Bravais lattice would intersect any of vertices, face mid-points or edge mid-points of the corresponding polyhedron as shown in  Fig.\,\ref{fig:bravais_1}. This observation was found to be extremely relevant for finding out the directions of basis vectors.

Following the idea, our next task was to detect the face and edge mid-points from the vertices of the polyhedron. This was accomplished by using the \textit{convex-decomposition} technique implemented in the \textit{Coxeter} toolkit \cite{Ramasubramani2021}. The vectors connecting each of these points and the geometric center of the polyhedron at (0, 0, 0) gave rise to a set of non-unit vectors. The number of vectors in this set was denoted by $g$. Our aim was to relate these vectors distributed on the surface of the polyhedron, with the original basis vectors $\vec{a}$, $\vec{b}$, $\vec{c}$ of the crystal. But, it was not possible to calculate the modulus of the basis vectors $|\vec{a}|$, $|\vec{b}|$, $|\vec{c}|$ without the detection of the complete unit cell. Towards that goal, in the next step, we aimed to detect the proper choices of the basis vector directions from the above geometric constructions. For this purpose, we considered all $\Comb{g}{3}$ combinations of vectors, a triplet was denoted by $\vec{a}_d$, $\vec{b}_d$, $\vec{c}_d$. Then, each triplet was chosen one by one and those triplets satisfying the previously detected crystal class were separated out into a set $\mathbb{T}_{cl}$. For this purpose, the isotropy of the moduli of the vectors, i.e.\,$|\vec{a}_d|$, $|\vec{b}_d|$, $|\vec{c}_d|$ was measured, and their mutual angles were calculated. These data were used to make decisions by following the conditions laid out in the Section 1 of Supplemental Material. It should be noted that the set of triplets, each consisting of three vectors, $\vec{a}_d$, $\vec{b}_d$, $\vec{c}_d$ was not the actual basis vectors but the directions of these vectors could always be detected by isolating the triplets for the respective crystal class. To handle real data with noise, two tolerance values; distance tolerance $\mathcal{X}_d$ and angular tolerance $\mathcal{X}_a$, were introduced while matching the conditions defined for the respective crystal class. All other triplets could be ignored for the simplicity of the calculations as those were no longer required. Such consideration allowed to handle the non-uniqueness of unit cell across multiple classes.
	
In general, the set $\mathbb{T}_{cl}$ could consist of multiple triplets corresponding to the same crystal class where each one was capable of producing the entire lattice by translation only, but this did not guarantee that each would satisfy the valid Bravais unit cell with allowed lattice sites. So, our next step was to check whether each triplet of the set $\mathbb{T}_{cl}$ would indicate a valid Bravais unit cell or not. We would like to reiterate that, the set $\mathbb{T}_{cl}$, which was constructed from the coordination polyhedron, provided only the directions of the basis vectors and not the lattice parameters themselves. The valid triplets consistent with the crystal class were added to another set, $\mathbb{T}_{latt}$, which was used for the next step.

\begin{figure*}
	\centering 
	\includegraphics[scale=0.4]{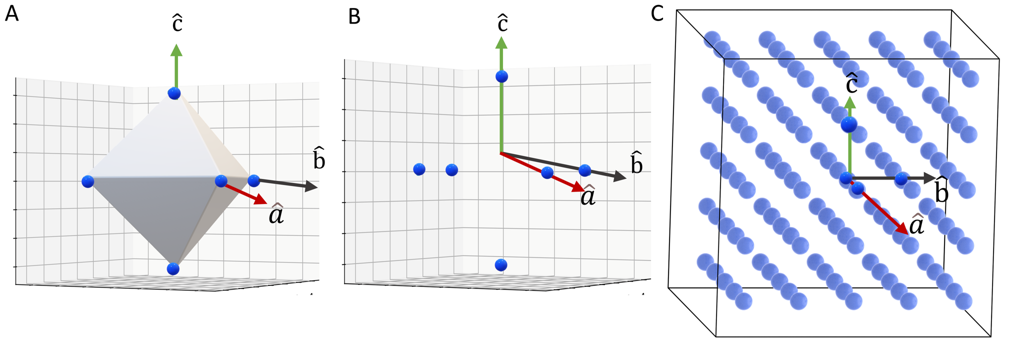}
	\caption{\textbf{The directions of lattice vectors are shown in SC Bravais lattice for validation purpose.} \textbf{(A)} The direction of the lattice vectors $\hat{a}$, $\hat{b}$ and $\hat{c}$, passing through the vertices of the constructed Octahedron are shown. The coordinate of the neighbors without the Octahedron along with $\hat{a}$, $\hat{b}$ and $\hat{c}$ as shown in \textbf{(B)}, indicates the direction of lattice vectors as the valid choice of unit cell for the Simple Cubic Bravais lattice, displayed in \textbf{(C)}. }
	\label{fig:ohvalidation}
\end{figure*}

For the validation purpose in the ideal SC crystal, we observed only one triplet satisfying the cubic class, which was chosen as the directions of basis vectors by default. Those directions are shown in Fig.\,\ref{fig:ohvalidation}A with the constructed Octahedron using the centroids of the clusters in the BOO. These centroids are also displayed in Fig.\,\ref{fig:ohvalidation}B without the polyhedron indicating the directions of basis vectors passing through the vertices of the constructed geometry. The data provided in Fig.\,\ref{fig:ohvalidation}C indicates the validation of the choice of basis vectors to replicate the entire SC structure.

It was important to mention, the choices of basis vectors directions did not depend on the fact whether these directions were commensurate with the simulation box vectors or not. Our method detected the global broken symmetry in terms of relative coordinates of the particles, i.e.\,the bond vectors, hence the relative orientation of the crystal with the simulation box was irrelevant. Even though for this example and in the synthetic data presented in Fig.\,\ref{fig:bravais_1}, this was the case, but in general for a simulated system, crystals seldom lined up with the simulation box. But the applicability of this algorithmic recipe remained unaffected. One such example is discussed below as in the Section \ref{sec:Results}.

\subsection{Determination of the triplets satisfying valid Bravais unit cells} \label{sec:alg_bravais}

We employed another search procedure using the all possible directional vectors of the set $\mathbb{T}_{latt}$.  In this section, our target was to isolate the triplets which satisfied the lattice sites of the allowed Bravais unit cells of the detected crystal class. This step could be further decomposed into three sub-steps, described below.

\subsubsection{Construction of the primitive unit cells} \label{sec:primitive_cell}
\begin{figure*}[!h]
	\centering 
	\includegraphics[scale=0.23]{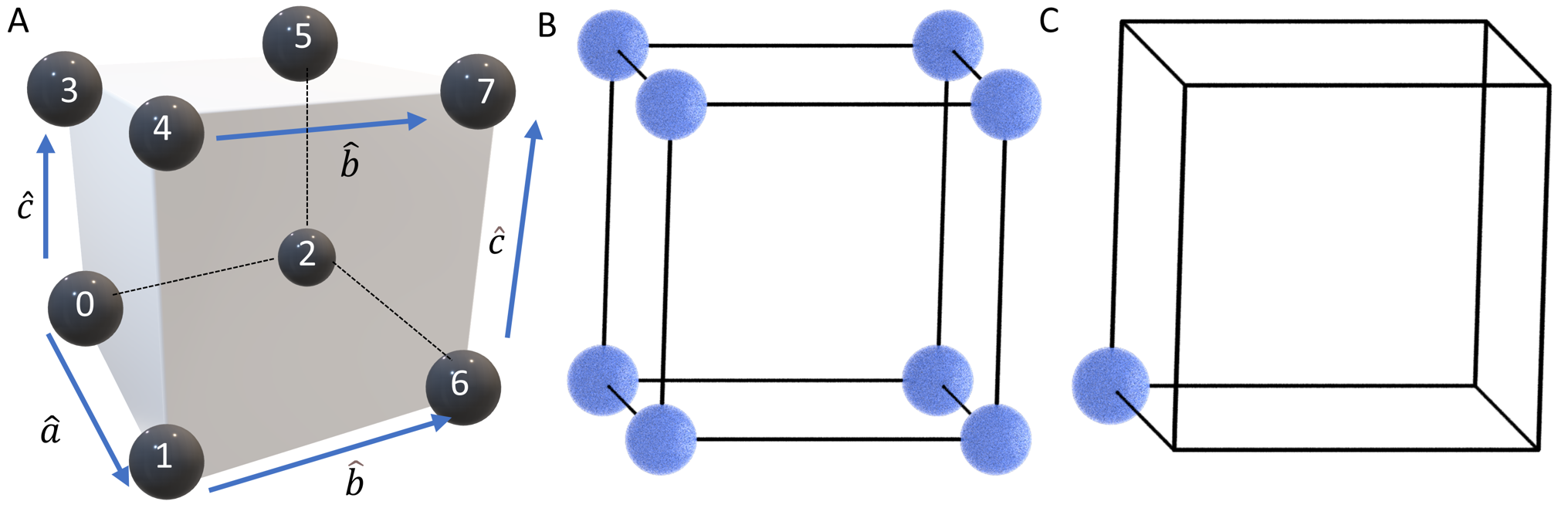}
	\caption{\textbf{The detected unit cell and equivalent particles are shown.}  The corner particles of the unit cell translated along the directions of basis lattice vectors are shown in a cartoon diagram in panel \textbf{(A)}. The unit cell of the SC lattice is displayed in panel \textbf{(B)} with the effective particles shown in panel \textbf{(C)}.}
	\label{fig:cartoon}
\end{figure*}

The unit cell of any Bravais lattice could be thought as a parallelepiped with the particles sitting at either the corners, face mid-points or at the center, including the primitive Hexagonal lattice. We first determined the parallelepiped from the particle coordinates and the possible lattice vectors  were extracted in the last step. A particle from the system was chosen at random, referred to as the reference particle $P_0$, and then for each triplet in $\mathbb{T}_{latt}$, the following search procedure was executed. From $P_0$, three other particles, $P_1$, $P_2$ and $P_3$ were determined by searching along $\hat{a}_d$, $\hat{b}_d$, $\hat{c}_d$ directions in the chosen triplet. These particles were the closest from the reference particle in the respective directions. These coordinates were real values from the input set of coordinates in the simulation reference frame. This protocol could be repeated from the newly found particles and rest of the corner particles of the parallelepiped could be easily figured out as shown in the Fig.\,\ref{fig:cartoon}A. Now we had the primitive unit cell in terms of real particle coordinates and from this information the true basis vectors for the chosen triplet of directions could be determined using Equations \ref{eq:eq_1}, \ref{eq:eq_2}. 

\begin{equation}\label{eq:eq_1}
	\vec{a} = \vec{r}_1 - \vec{r}_0 \text{ \& } \vec{b} = \vec{r}_2 - \vec{r}_0 \text{ \& } \vec{c} = \vec{r}_3 - \vec{r}_0 
\end{equation}
\begin{equation}\label{eq:eq_2}
	\alpha = \cos^{-1}(\hat{b} \cdot \hat{c}) \text{ \& } \beta = \cos^{-1}(\hat{a} \cdot \hat{c}) \text{ \& } \gamma = \cos^{-1}(\hat{a} \cdot \hat{b})
\end{equation}
where $\vec{r}_0$, $\vec{r}_1$, $\vec{r}_2$ and $\vec{r}_3$ are the coordinates of the particles $P_0$, $P_1$, $P_2$ and $P_3$ in the global reference frame respectively. With real data, distance and angular cutoffs were used to conduct the searches.

It was important to note that though the basis vectors were obtained by tracking the real particles, we still did not have the information whether these triplets successfully satisfied the valid Bravais unit cells or not. It could be done once the lattice parameters as well as coordinates of lattice sites were compared with the standard crystallographic convention as long as we initially considered one particle per lattice site.

\subsubsection{Construction of the conventional unit cells} \label{sec:conventional_cell}
The next order of business was to find the conventional unit cells for all these choices in $\mathbb{T}_{latt}$. It involved finding the positions of the particles sitting at the face mid-points or edge mid-points of the primitive unit cells as the Bravais lattice was considered as the system of interest. We employed a technique based on construction of the convex hull from the corner particles of the parallelepiped. The \textit{Delaunay} module in \textit{Scipy} \cite{2020SciPy-NMeth} was used for this purpose. Following this method, we obtained the particle coordinates staying either inside of the parallelepiped or on the face of the parallelepiped except the corner ones. To handle the real data with noise, the hull was expanded isotropically by a volume factor ($\mathcal{V}$), so that all particles located at any sites other than the corners, could be taken into account. For example, if the system corresponded to a face-centered lattice, the particles exactly located on the faces of parallelepiped could be missed due to the noise in case the hull was not expanded slightly; as a result one would end up with wrong detection of the unit cell. To handle these circumstances, a little bit expansion of the convex hull was needed which required human intervention and the value of $\mathcal{V}$ strictly depended on the level of noise. For an ideal crystal structure, $\mathcal{V}$ was set to 0.

For the example of SC crystal, no other particles were found on the faces or inside the parallelepiped as the SC unit cell was primitive in nature. This was checked by searching the system particles whether the convex hull included any other particles except the corner ones, as a result, this particular step added no further information to the existing data as shown in Fig.\,\ref{fig:cartoon}B. But in general, this step was necessary to perform in order to identify any conventional unit cell if it existed in the system.

\subsubsection{Coordinates of effective particles in each of the conventional unit cells}\label{sec:effective_particles}
After identifying the dimensions of a possible unit cell, the only remaining aspect for the full specification of the entire unit cell was the effective number of particles in it. We would like to point out, that at this stage, we still did not know what combinations in $\mathbb{T}_{latt}$ were crystallographically viable. To make that final decision we need the effective number of particles along with the lattice parameters. This will be discussed in this step.

Two particle with coordinates, $\vec{r}_1$, $\vec{r}_2$, were considered as non-equivalent, if the joining vector between them, $\vec{r}_{12} = \vec{r}_2 - \vec{r}_1$ did not coincide with any of the basis vector directions i.e., $\hat{a}$, $\hat{b}$, $\hat{c}$ and the modulus of $\vec{r}_{12}$ i.e.\,$|\vec{r}_{12}|$ sufficiently differed from the any one of the lattice dimensions, $|\vec{a}|$, $|\vec{b}|$, $|\vec{c}|$. For actual calculations, distance and angle tolerances were employed. If it did coincide, the coordinate of one particle could be generated from the other one by translating along the respective basis vector. Hence, any one particle could be considered as the effective particle. If the condition was violated both the particles would be crystallographically distinct and needed to be considered as effective particles. The algorithmic way towards the determination of effective particle coordinates is shown in the section 3 of Supplemental Material.

This procedure was repeated for each triplet in the set $\mathbb{T}_{latt}$ and the number of effective particles and their lattice sites were monitored and those triplets were isolated satisfying allowed Bravais unit cells under the detected crystal class. It should be noted that as we considered single crystalline structure initially, it was quite certain to obtain only one type of Bravais lattice upon completion of this protocol. So, the fact of obtaining multiple types of Bravais unit cells within the same crystal class could be discarded if this step was executed properly.

In the example of SC crystal, only one triplet was found in the set $\mathbb{T}_{brav}$ satisfying the cubic crystal class. The basis vectors appeared to be $\vec{a}$ = [6.315, 0, 0], $\vec{b}$ = [0, 6.315, 0], $\vec{c}$ = [0, 0, 6.315] which were estimated from the coordinates of eight corner particles using Equation \ref{eq:eq_1}. The lattice parameters corresponding to the triplet were $a$=6.315, $b$=6.315, $c$=6.315, $\alpha$=90$^{\circ}$, $\beta$=90$^{\circ}$, $\gamma$=90$^{\circ}$ suggesting the exact matching of the detected unit cell with the initially considered one which was used to replicate the entire lattice.

\begin{figure*}
	\centering 
	\includegraphics[scale=0.23]{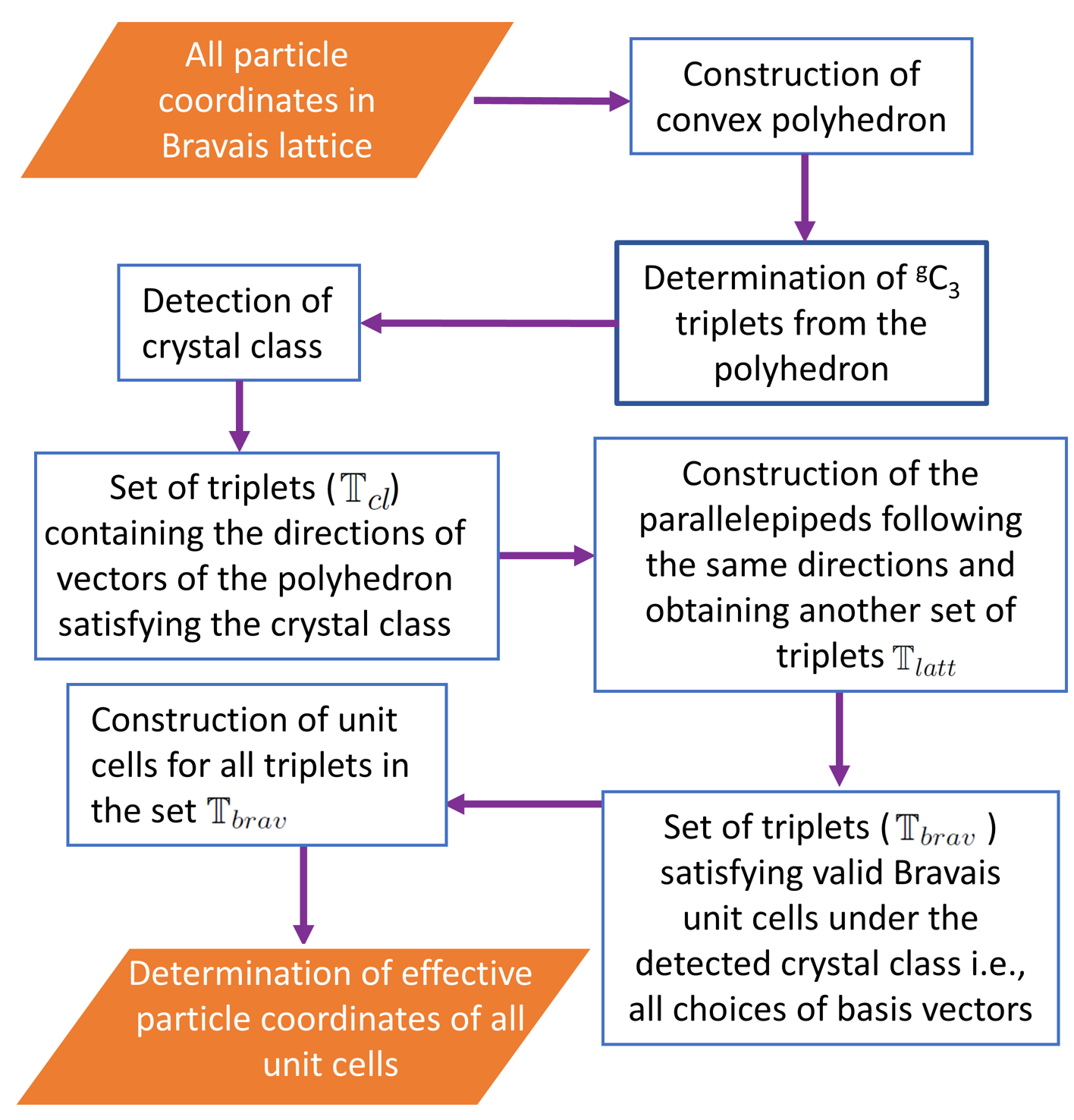}
	\caption{A flowchart is shown illustrating the technical steps to detect all valid unit cells starting from the Bravais lattice}
	\label{fig:bravais_direc}
\end{figure*}

\section{Determination of unit cell for a general multi-component system}\label{sec:alg_nonbrav}
The straightforward application of the above algorithm to a general system was not possible for two reasons. An arbitrary system can have multiple types of particles, and it was not possible to determine, in an \textsl{a priori} fashion, how those different types contributed to the basis. Even for a single component system, predicting the basis from particle coordinates alone, without further considerations was not possible. In the absence of this information, associating particle coordinates with lattice sites could not be done, which was an essential feature of the algorithm used for a simple Bravais system described in the last section. Apart from this, the cornerstone of our method was the unique geometric construction of the polyhedron from the BOO diagram. Using the polyhedron to search for possible lattice vector directions worked because of the existence of a single local environment in a Bravais system. For a general crystal, where the associating some particle with the underlying lattice sites would not, in general, give rise to a simple Bravais description and will result in multiple types of local environments. This fact was described in the beginning in our discussion of the concept of local environments. Under such setting all the steps that led us to the unit cell parameter would not work, rendering the straightforward application of the heuristics steps useless.

We solved this problem in two steps. First, we separate the system based on species labels,  in case of a multi-component system. This is extremely crucial, a failure to do so can easily break down the heuristic ideas that use direct particle coordinate data. This step in described next.

After filtering out one type of particles, there was no guarantee that the resulting system was a simple Bravais crystal with a single particle basis. However, this was not a fundamental bottleneck, as any non-Bravais lattice can be systematically transformed into a simple Bravais description (of single type) by another set of conceptual steps that respected crystallographic restrictions. We exploited this observation and constructed a transformed Bravais system from the first round of transformed system consisting of a single type of particle. The transformed Bravais system could be anlalyzed by the already described algorithm to find out the crystal class. And from there, the lattice parameter and equivalent positions of the original crystal could be determined.

The steps, along with the proper justifications and possible conceptual pitfalls, are described in the rest of the section. For demonstration purposes, a perfect crystallite constructed from the experimental parameters of the $\beta$-Mn crystal, which has a space group of $P4_{1}32$ and two Wyckoff positions. We chose this example for its sufficient complexity, it has $20$ particles in a unit cell \cite{Xie2013}, which makes it extremely difficult to elucidate the structure from a set of particle coordinates by visual inspection. For this structure, all particles type appeared to be similar; hence the identity separation technique was not required to perform. Output of each step is presented in the relevant subsection. The appropriate consideration for statistical noise is discussed wherever it is relevant without showing actual data, during the discussion. These issues are dealt with real examples from Monte Carlo simulations discussed in Section \ref{sec:Results}.

\subsection{Separation of the system based on species labels of the particles}
In a single crystalline material with multi-particles basis, the translational symmetry depends on the periodicity of the whole basis. In in the interest of of our current context we repeat some standard facts about lattices and basis in peridoic arrangements in crystals. Complete specification of the whole crystal may not be obvious if one focuses on a particles of one type of species label, if proper care is not exercised. Another fact about general crystals is the some particles can be associated with certain sites of the underlying lattice without loss of generality. In doing so, certain lattice sites can remain unoccupied by real particles. None of these changes the overall nature of the crystal. Following this argument, it is straightforward to conclude that by focusing on the species label that is sitting at the lattice sites, one can figure out the details of the unit cells using heuristic algorithms presented above. The details are not obvious because of the possible non-Bravais nature of the lattice, but those can be dealt with as we describe later.

Disregarding the species label could lead to erroneous conclusion while relying on the kind of heuristics our method is based on. This point can easily be understood in the well known rock salt structure with inter-penetrating FCC lattices of \ch{Na+} and \ch{Cl-} ions.  If the species labels were ignored and one were to focus on the particle coordinates alone, the structure would appear as a simple cubic crystal with lattice constant ``$a/2$'', $a$ being the lattice parameter of the original structure. This fact is illustrated in Fig.\,\ref{fig:nacl}. Hence, any particular atom type must be separated out, in order to satisfy the periodicity of the basis (both \ch{Na} and \ch{Cl} atoms) in a crystal. But, associating a vertex of the corresponding unit cell to either a \ch{Na} or \ch{Cl} atom does not affect the symmetry of respective crystal structure. Here, the unit cell is translationally shifted only, where the periodicity of basis takes care of the overall crystallographic symmetry.

In the light of this discussion, the first step of our algorithm for general systems was to pick a species label in the input coordinates and isolate all particles with that label. This reduced system was used for analysis based on particle coordinates alone. This did not affect the structure recognition as all types of particles were expected to be present in the basis to define the full crystalline symmetry. Therefore, anisotropic positional distribution for any particular type of particles at a cutoff distance, would encompass essential features of the broken symmetries and provide a good starting point for our algorithm in the general case. In other words, coordinates of any one species label would be sufficient for partial realization of our ideas laid out before. Fo detecting the full unit cell, one would eventually need to incorporate all coordinates, irrespective of species label, as we will show later in the paper. The anisotropic positional distribution of local neighbors would vary with particle type, but our heuristic would guarantee the same lattice parameters could be extracted from a single crystalline material. This fact is presented as a cartoon diagram in Fig.\,\ref{fig:nacl}A. This discussion confirmed that if no particular particles type produced anisotropic distribution then recognition of crystal structure was not possible at all, hence the input data should be questioned. After separating the particles of similar identity, we could consider the respective coordinates only where the identity information were ignored to perform the next steps. We would like to emphasize, for a system with the particles having similar identity, this step was not required to perform; only coordinate data were enough to proceed further with the algorithm. An example of \ch{NaCl} structure was used to illustrate the recipe as shown in Fig.\,\ref{fig:nacl}B a, b where the entire system is classified into two; one for \ch{Na} atom and other for \ch{Cl} atom.

The $\beta$-Mn example, being a single component system, used for illustration purpose does not have this complexity.  We will illustrate this point of our algorithm in one of the test systems presented in the section 8.1 of Supplemental Material.

\subsection{Detection of multiple local environments}\label{sec:alg_detect_bravais}
After filtering one type of species label, the resultant reduced system was the main focus for the current step. This is not true for the entirety of the algorithm, which would conclude by considering all the particles in the input system. But for the sake of heuristics we would be dealing with reduced systems and that point will be made clear at appropriate places.

{The decision that needed to be made was whether the reduced system was a simple Bravais lattice or not. It was easy to detect algorithmically following the ideas of local environments introduced in Section \ref{sec:local_env}. As mentioned before, a Brvais system, with one particle basis, had all particles with identical environments. This idea was used for identification of the Bravais nature of the reduced single component system. The first step was to construct the bond order diagram with a suitable cutoff distance $r_c$. All the remarks made about an appropriate value of cutoff in the last section applied here. The clusters of points in the BOO, were denoted by $\mathcal{C}_1, \ldots, \mathcal{C}_{N_{c}}$. Here, each $\mathcal{C}_{\alpha}$, with $\alpha=1,2,\ldots,N_{c}$, was identified by the K-means method as mentioned before. Each $\mathcal{C}_{\alpha}$ represented a unique value of the pairwise bond vector depicted in the BOO diagram, where the center was at (0,0,0). In other words, the local environment of $i$-th particle, $\mathbb{G}_{i,r_c}$, which was a set of bond vectors with its neighbors, could also be defined as a set of cluster labels for the respective bond vectors. For the sake of convenience, constructed the set $\mathbb{E}_{i,r_c} = \{{\alpha}, {\beta},\ldots\}$, with $\alpha \ne \beta = \ldots = 1,2,\ldots,N_c$, for the $i$th particle. This representation of local environments, i.e.\,in terms of $\mathbb{E}_{i,r_c}$s instead of $\mathbb{G}_{i,r_c}$s, was completely equivalent and similarity of environments of two particles could be ascertained by comparing the cluster indices belonging to these two sets. The mapping between the labels of clusters and the actual bond vectors were straightforward and could be determined from the Cartesian coordinates of the $N$ particles followed by the clustering. 

For an ideal system, the centroid of the clusters were not required to be calculated separately due to the absence of any positional fluctuations leading to the coordinate of the centroid of a cluster coinciding with the points. But, in real simulation data, the identification of the clusters was important followed by the determination of the centroid of the clusters. }The number of clusters $N_c$ explicitly depended on the complexity of the crystal and the chosen value of $r_c$. Despite having a complex crystal, a sufficiently large value of $r_c$ could slow the convergence of the algorithm. It was important to choose the correct value which required human intervention. In this case, given the positions of the centroids $\mathcal{C}_{\alpha}$'s, a particular bond vector between neighboring particles $i$ and $j$ could be identified by a particular cluster label within a tolerance. Following this simple procedure the environments can be enumerated for all particles, $\{\mathbb{E}_{1,r_c}, \mathbb{E}_{2,r_c},\ldots, \mathbb{E}_{N,r_c}\}$. 

After detecting the local environment of all particles, the next step was to isolate the subset of particles with identical environments. The number of such particles were denoted by $N_s$, for a fixed value of the cutoff $r_c$. In general $N_s < N$, but in case of the Bravais lattice with one particle per lattice site, by definition all particles have the same environments, resulting in $N_s = N$. For simulation output with noise, exact equality was replaced by approximate one within a tolerance. Here, to make the decision, it was more convenient to check for the number of clusters as well, which was equal to the coordination number, $N_c = N_{CN}$. The initial environment detection on the model  $\beta$-Mn crystal system with $r_c=2.6$ gave $N_s = 0.05 \times N$, indicating the system was not a Bravais lattice with one particle basis.

\begin{figure}[!h]
	\centering 
	\includegraphics[scale=1.0]{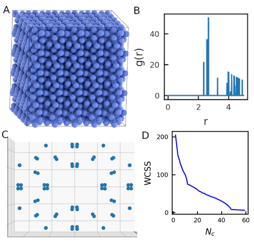}
	\caption{\textbf{Local positional environments and detection of clusters are shown for $\beta$-Mn crystal.} An ideal system of $\beta$-Mn crystal is shown in panel \textbf{(A)} with the corresponding RDF shown in panel \textbf{(B)}. Detecting the nearest neighbor distance $r_c$ $\sim$ 2.6, the BOO is displayed in panel \textbf{(C)} indicating the existence of forty eight clusters of particles as confirmed by the \textit{K-Means} algorithm as shown in \textbf{(D)}.}
	\label{fig:nonbrav_analyses}
\end{figure}

\begin{figure}[!h]
	\centering 
	\includegraphics[scale=1.0]{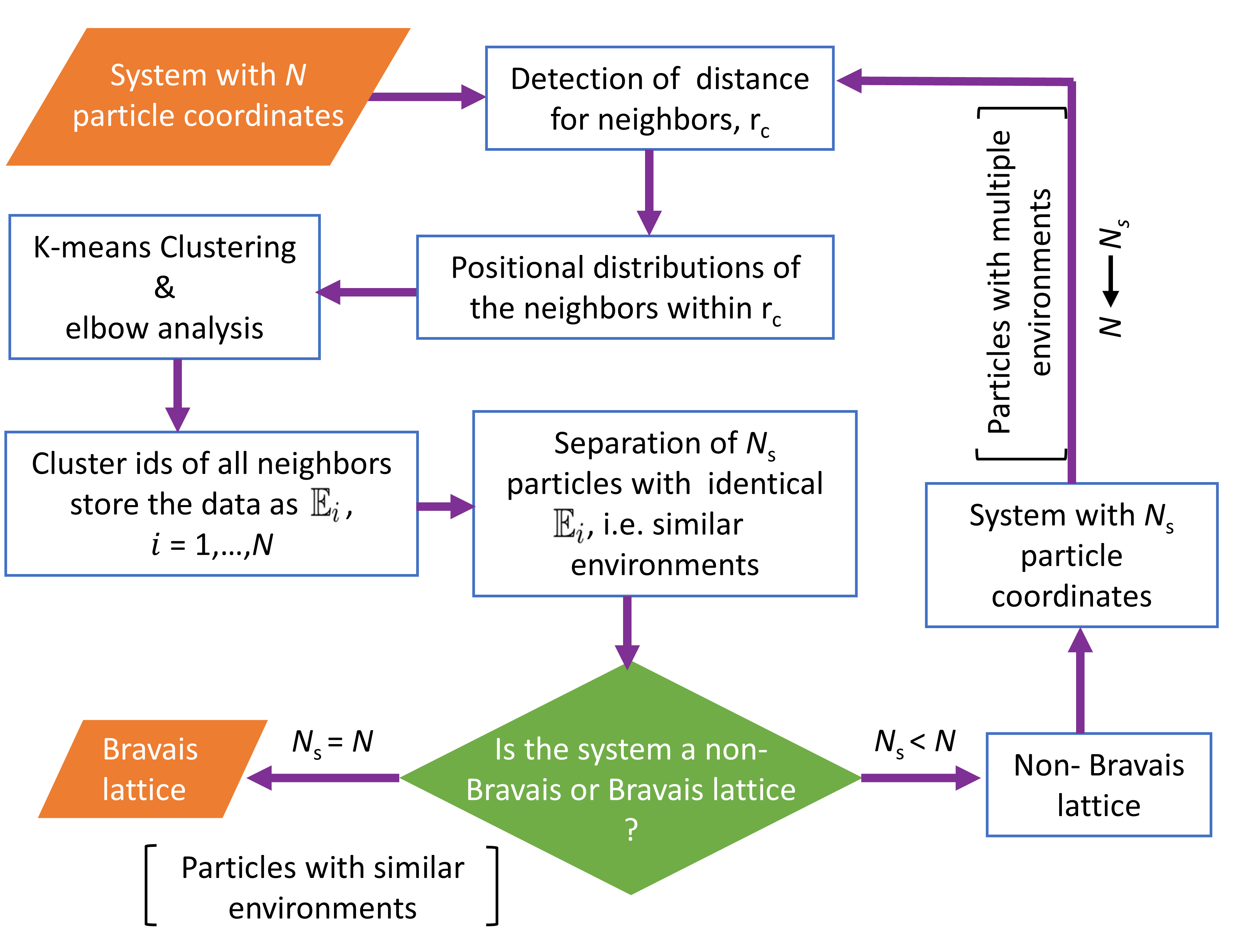}
	\caption{The schematic diagram of the ``environment separation'' technique is shown}
	\label{fig:envsep}
\end{figure}

\subsection{Separation of local environments}\label{sec:alg_env_sep}
The final goal of this important step was to obtain a subset of coordinates with the same species label, from the original set, which formed a Bravais lattice with one particle at each lattice site. These positions were unaltered from the initial coordinate set but shared this important characteristics. The advantage of this set-up would be evident later in the next major steps towards final detection of the unit cell of original non-Bravais lattice by appending the similar protocol developed for Bravais lattice. These subsequent steps were performed for a set of coordinates that had multiple local environments, i.e.\, a non-Bravais lattice only. With this remark, we describe the environment separation algorithm in this sub-part.

The key insight here was to appreciate that the local environment was a function of the cutoff distance used to define the neighbors. If one looked at the subset of $N_s$ particles, while ignoring the rest, and repeated the environment detection as described in the previous step, then in general, there was no guarantee that there would be a single type of local environment. In this step, the original coordinates of $N_s$ particles were considered and the BOO diagram was calculated with an appropriate value of $r_c$, ensuring all the restrictions mentioned before. If the new subsystem was still non-Brvais in nature the entire procedure was repeated after filtering out particles with identical local environments. It was found that after a finite number of iterations, eventually there would be a set of particle coordinates left that formed a simple Bravais lattice. The number of iterations depended on the complexity in the crystal and chosen value of cutoff distance $r_c$ ensuring the iteration to be continued until all the particles in the transformed system had similar environments. In summary, this iterative procedure filtered out certain particles, while changing the length scale for defining the neighbor, until the final objective was attained. The flowchart up to this step is displayed in Fig.\,\ref{fig:envsep} with the algorithm presented in the supplemental information.

\begin{figure*}[!h]
	\centering 
	\includegraphics[scale=0.4]{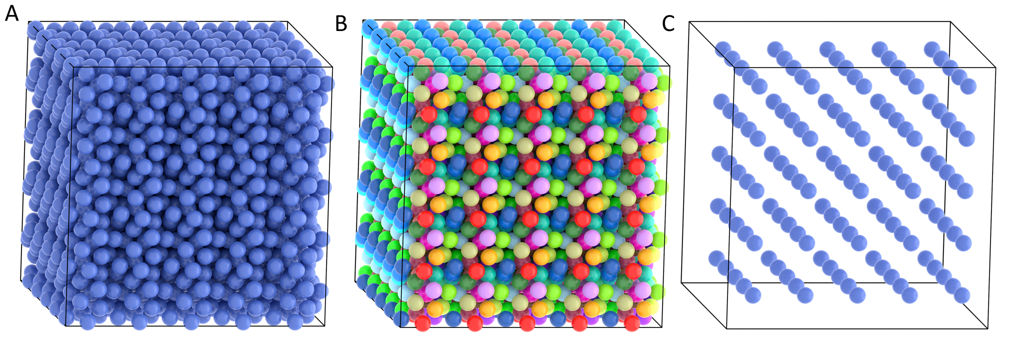}
	\caption{\textbf{Environment separation of $\beta$-Mn structure is shown in three steps:} \textbf{(A)} An ideal system of $\beta$-Mn crystal, \textbf{(B)} the same system with atoms in multiple colors based on the similar kind of environment and \textbf{(C)} a reduced Bravais lattice after environment separation are shown.}
	\label{fig:bravais}
\end{figure*}

For the example of $\beta$-Mn crystal, the environment separation was implemented as it appeared to be non-Bravais. The particles were categorized based on the local environments with the cutoff value $r_c$ set at 2.6. The configuration is shown in Fig.\,\ref{fig:bravais}B with the particles displayed in multiple colors where the particles in identical color had similar environment. The final output of the environment separation is shown in Fig.\,\ref{fig:bravais}C. This reduced system contained 5\% of the total number of particles in the original system. The local positional environment of this system was analyzed again which confirmed that the transformed system was a Bravais lattice. In this case, only one iteration was required to separate the particles with similar environment. It was important to note, the isolated particles shown in Fig.\,\ref{fig:bravais}C were not generated artificially, instead, these particles were extracted from the original system based on the special characteristics of having the same local environments.

\subsection{Determination of the crystal class and basis vectors from the analysis of the transformed Bravais lattice}
\begin{figure*}
	\centering 
	\includegraphics[scale=0.4]{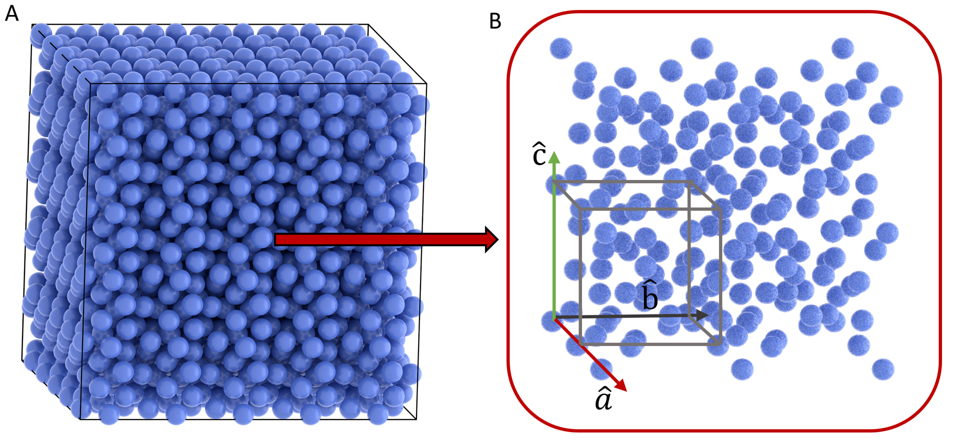}
	\caption{\textbf{Validation of the directions of lattice vectors are shown in a chunk of $\beta$-Mn crystal.} \textbf{(A)} The non-Bravais lattice system with $N$ particle coordinates are shown. \textbf{(B)} A random chunk of particles from the system are chosen and the direction of the lattice vectors, $\hat{a}$, $\hat{b}$ and $\hat{c}$ are shown along with the unit cell for the validation purpose.}
	\label{fig:nonbravaisvalidation}
\end{figure*}

After converting the non-Bravais lattice into a Bravais one, the stage was set to use the existing algorithm as discussed in the Section \ref{sec:brav_alg}. It was important to recall those steps again which led to the detection of crystal class and directions of basis vectors considering the Bravais lattice. The set of all possible triplets of vectors, consistent with the detected crystal class was referred to as $\mathbb{T}_{brav}$. For the system of $\beta$-Mn crystal, the transformed Bravais lattice appeared to be SC crystal in nature which could be even confirmed by the visual inspection only as shown in Fig.\,\ref{fig:bravais}C. The previous approach applied on the SC Bravais lattice, confirmed the cubic class with the basis vectors $\hat{a}$ = [1, 0, 0], $\hat{b}$ = [1, 0, 0], $\hat{c}$ = [0, 0, 1]. These vectors agreed well with the initial consideration (see Fig.\,\ref{fig:nonbravaisvalidation}) and were justified to be the valid choice for the corresponding Bravais lattice as suggested by the Fig.\,\ref{fig:ohvalidation}C.

\begin{figure*}[!h]
	\centering 
	\includegraphics[scale=0.85]{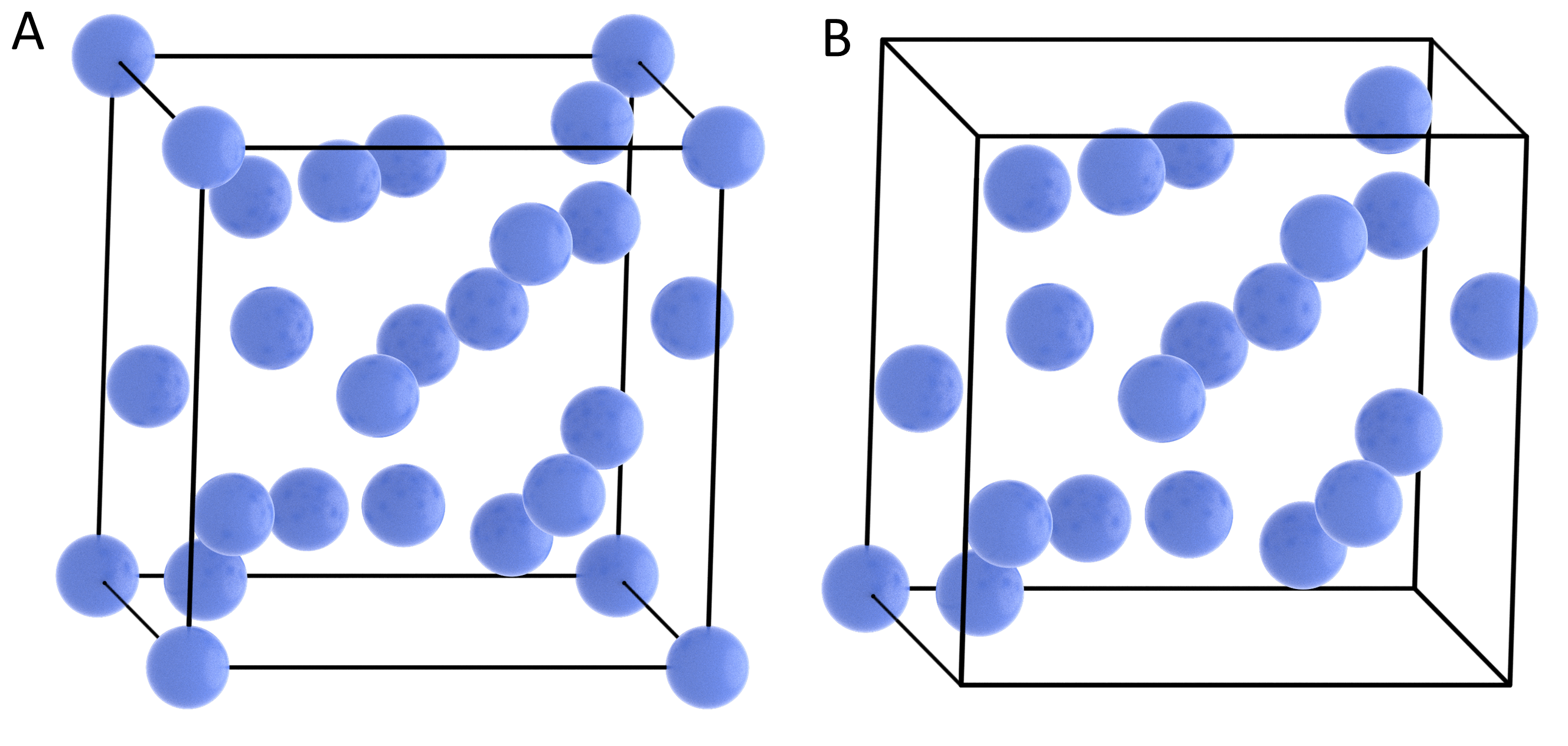}
	\caption{\textbf{The unit cell of $\beta$-Mn crystal is shown.} All particles in the unit cell are displayed in panel \textbf{(A)} and effective particle coordinates are shown in panel \textbf{(B)}. There exists 20 effective coordinates in the unit cell of $\beta$-Mn crystal.}
	\label{fig:betamn_uc}
\end{figure*}

\subsection{Construction of all valid unit cells of the original system} \label{sec:alg_uc_original} 
The lattice parameters detected from the Bravais system as described in Section \ref{sec:alg_bravais}, i.e.\,corresponding to the triplets in $\mathbb{T}_{brav}$, were identical to that of original system. This meant, the detected parallelepipeds from the corresponding Bravais systems served as the envelopes for the valid unit cells of the original non-Bravais system with exact basis vectors under the same crystal class. Our remaining job was to find out the coordinates of the rest of particles contributing to the unit cells of the original system.

We considered all particle coordinates in the original system irrespective of single or multiple particle types and for each triplet in the set $\mathbb{T}_{brav}$, the similar strategy was followed to search for the particles staying inside or on the surface of the respective parallelepipeds. The exact protocol was already described in the second paragraph (b) of Section \ref{sec:alg_bravais}. In this way, we obtained all the particle coordinates contributing in the corresponding parallelepipeds for all triplets, all of which served as the valid choices of unit cells of the original structure. It was important to remember, in this step all particle coordinates irrespective of types were taken into account to construct the unit cells. The effective particle coordinates in the unit cells were also evaluated following the method discussed in the paragraph (c) of Section \ref{sec:alg_bravais}. It confirmed that the unit cell of $\beta$-Mn crystal structure consisted of 20 effective particle coordinates \cite{Xie2013} with lattice parameters $a$=6.315, $b$=6.315, $c$=6.315, $\alpha$=90$^{\circ}$, $\beta$=90$^{\circ}$, $\gamma$=90$^{\circ}$ as shown in Fig.\,\ref{fig:betamn_uc}B. It showed that our proposed algorithm was able to detect the exact crystal class, all valid choices of basis vectors, corresponding lattice parameters and so the respective unit cells for any crystal structure irrespective of simple and complex one from the particle coordinates only.

Following the prescribed protocol, we detected all the valid choices of basis vectors and respective effective coordinates of every single valid unit cell (as described in the section 3 of Supplemental Material). In real system with noise, it would create a problem if the randomly detected unit cell appeared to be sufficiently defective. In such scenario, the followed path should include the determination of multiple unit cells using this algorithm and consideration of the average values of lattice parameters and respective fractional coordinates.

\begin{figure*}[!h]
	\centering 
	\includegraphics[scale=0.23]{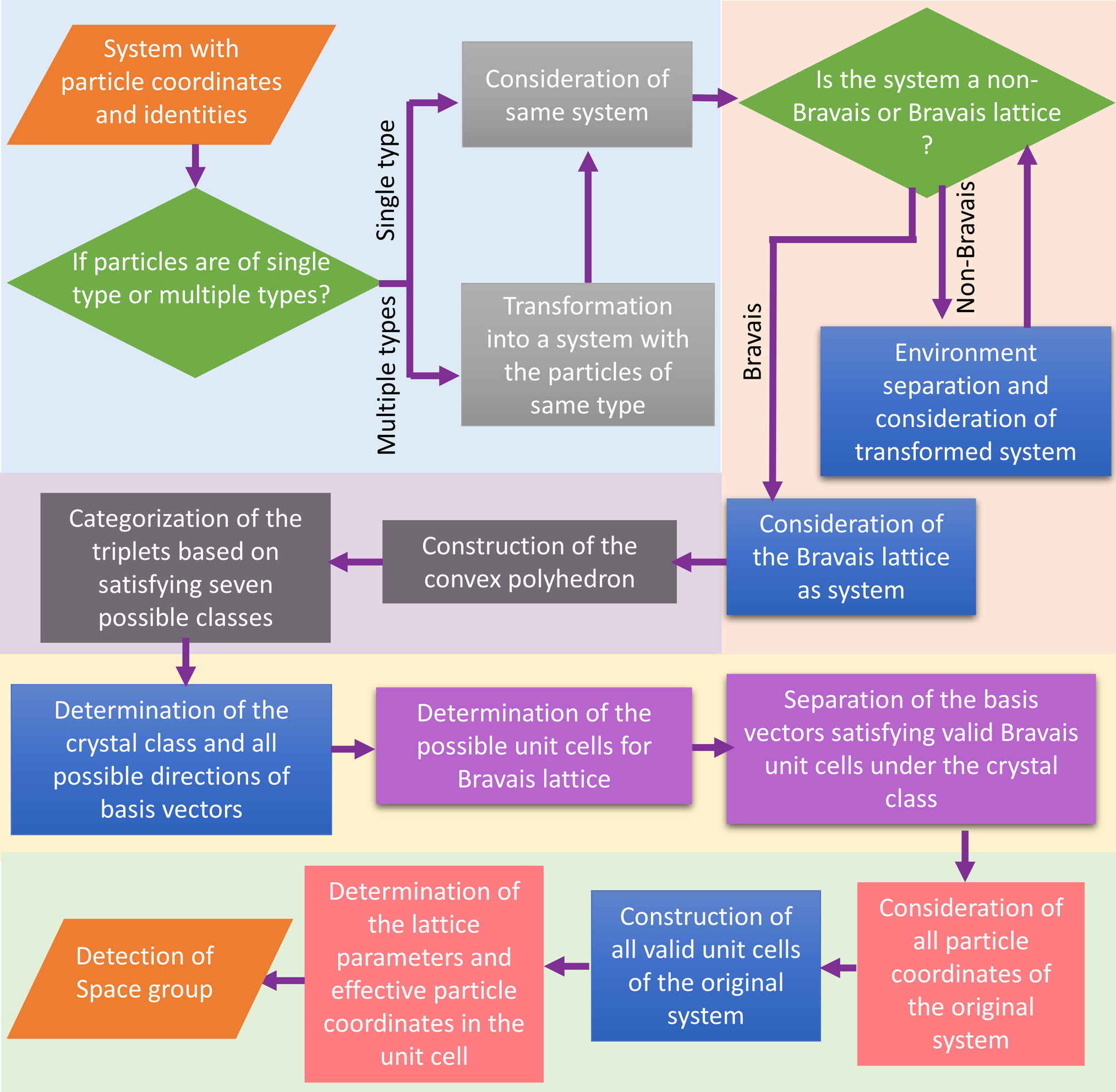}
	\caption{The complete algorithmic approach is illustrated in this figure.}
	\label{fig:summary}
\end{figure*}

\section{Determination of space group of the crystal structure} \label{sec:alg_space_grp}
As the final goal, we wanted to determine the space group of the crystal structure. For this, \textit{Spglib} package \cite{spglibv1} was used to operate all the possible crystallographic symmetry operations on the crystal structure, which provided the space group of the crystal as output. While performing this step, we had complete information of the unit cells i.e., basis vectors and coordinates of the effective particles for any simple Bravais or non-Bravais lattice system. The fractional coordinates of the particles were estimated for each unit cell following standard method which was standard in the field of crystallography and fed into the \textit{Spglib} program as inputs. This piece of code checked all the valid symmetry operations for the unit cell barring two tolerances, the distance tolerance ``\textsl{symprec}'' and angle tolerance ``\textsl{angle\_tolerance}''. For an ideal crystal, each of the tolerance values could be set at zero. As the output of this package, space group of the crystal structure was obtained according to the standard international nomenclature \cite{IuCr2016A}. All the symmetry operations could be possible to access using ``\textsl{get\_symmetry}'' module. It appeared that space group detection was quite straightforward using this package for an ideal crystal and it was complicated in case of a noisy crystal, as the tolerance values were required to be tuned properly. The same space group were determined for all  triplets of the set $\mathbb{T}_{brav}$. Multiple choices of the basis vectors as well as effective particle coordinates did not affect the space group because the fractional coordinates defining the crystalline symmetry remained unique for all cases. 

The space group of ideal $\beta$-Mn crystal appeared to be $P4_{1}32$ according to the international symbol. The complete flowchart of the algorithm is illustrated in Fig.\,\ref{fig:summary} and corresponding steps are showcased as a single figure in the supplemental information. In the next section, few examples will be discussed upon the implementation of the algorithm as the validation purposes. These results will mainly focus on the necessary steps of the algorithm in the context of the corresponding crystal structures.

\section{Results} \label{sec:Results}
While presenting our new heuristic algorithm for crystal structure detection from real space computer simulation data in the last section, we used a perfect crystal of $\beta$-Mn as a test case to demonstrate the working of each and every step. At the end of the entire process, we showed that the experimental parameters of the unit cell used to prepare the crystallite were reproduced by the algorithm, confirming the correctness of the overall method. This could be regarded as the demonstration of proof of principle in an idealized but fairly non-trivial system. The non-trivial aspect of the demonstration stemmed from the fact that $\beta$-Mn had 20 atoms in the unit cell, which was virtually impossible to detect from the full crystalline system by visual inspection, even for an experienced investigator. Despite this, several important aspects of the method, that could pose potential complications in realistic situations, were not verified. Controlled investigations to test a particular feature on carefully constructed test systems are presented in the Supplemently Information. In the first test case (section 7 of Supplemental Material), we show the effect of the simulated noise in the context of the same $\beta$-Mn system. A second validation, we deal with a multicomponent system - an ideal crystal of strontium oxalate (\ce{C3O6Sr}) - and illustrate how compositional complexity is handled by following the procedure already outlined in the previous section (section 8.1 of Supplemental Material). The data for two more cases: (i) a body-centered cubic (BCC) crystal structure of Barium with space group $Im\bar{3}m$ and a hexagonal-closed packed (HCP) structure of potassium sulfate (\ce{K2SO4}) with space group $P6_{3}/mmc$, are presented in the sections 8.2 and 8.3 of Supplemental Material, respectively. Together, they demonstrate the ability of the method to deal with all types of crystal structure.

In this section, we present the results of structure determination in a computational self-assembly problem. Broadly speaking, this example provides the complete test case for intended application of this method, howsoever, variants around the central idea are likely to have even wider relevance. A system of hard truncated tetrahedra a known to spontaneously form cubic diamond structure from the dense liquid phase in Monte Carlo simulations \cite{Zhang2005, Dshemuchadse2011, Wang2017, Damasceno2012, Klotsa2018, Marson2019} (for details, see section 5.1 and 5.2 of Supplemental Material). Three systems, corresponding to three packing fractions ($\phi$) in the solid part of the phase diagram with the same crystal structure, were subjected to structure determination. These three situations corresponded to increasing statical noise typical of simulation systems as the $\phi$ was decreased (see the BOO diagrams in Figs.\,\ref{fig:TT}A,B,C) for $\phi = 0.89, 0.7, 0.6 $, respectively. This exercise validated our method in the presence of real noise and other potential issues one is likely to find in the intended application of the method. The method was successful in determining the correct structure in all three situations, salient features of the intermediate steps are discussed below.

{The system being a single component one, did not require identity separation. For environment detection RDF based cutoffs worked and appropriate $r_c$ values increased with decreasing $\phi$, and the values of $\sim$ 1.1, $\sim$ 1.3 and $\sim$ 1.5 gave two types of local environments. These are shown as two colored particles in all three systems. It indicated that as long as the system is crystalline, noise did not destroy the local environments, which is the cornerstone of our method. In the next step the non-Bravais systems were put through the environment separation protocol and all of them took the same number of iteration to produce the corresponding transformed Bravais system with the same crystal structure, namely FCC. The $r_c$ values that ultimately gave the environment separation was set at $\sim$ 1.75, $\sim$ 1.9 and $\sim$ 2.0 respectively. A slightly higher value for the low density crystal was perfectly in line with the usual expectation.

The biggest difference in three step ups arose in the next step where all possible directions of the basis vectors and the actual basis unit cell dimensions were determined. The distance and angular cutoff, $\mathcal{X}_d$ and $\mathcal{X}_a$, respectively needed to be adjusted carefully, especially at the low $\phi$ system with the maximum noise. In the presence of very minimal noise at $\phi \sim$ 0.89, the choice of basis vectors was unique; $\vec{a}$=[-0.395,  0.212, -0.940], $\vec{b}$=[-0.534, 0.811, 0.383] and $\vec{c}$=[-0.791, -0.646,  0.216] and lattice parameters were $a$=2.101, $b$=2.107, $c$=2.16, $\alpha$=89.87$^{\circ}$, $\beta$=89.30$^{\circ}$, $\gamma$=88.88$^{\circ}$ within the tolerances, $\mathcal{X}_d$ and $\mathcal{X}_a$ set at 0.05 and 5$^\circ$ respectively. At the moderate noise level at $\phi$ $\sim$ 0.7, the directions of basis vectors appeared to be same with the lattice parameters, $a$=2.17, $b$=2.19, $c$=2.25, $\alpha$=89.17$^{\circ}$, $\beta$=88.56$^{\circ}$, $\gamma$=90.07$^{\circ}$ within the tolerances, $\mathcal{X}_{d}$=0.1, $\mathcal{X}_{a}$=10$^\circ$. Similar thing also happened for the system with sufficiently high noise level ($\phi \sim$ 0.6). In this case, the lattice parameters were $a$=2.26, $b$=2.28, $c$=2.36, $\alpha$=89.06$^{\circ}$, $\beta$=87.27$^{\circ}$, $\gamma$=92.15$^{\circ}$ barring the tolerances, $\mathcal{X}_{d}$=0.12, $\mathcal{X}_{a}$=10$^\circ$. The unit cells where all particles were shown rather than only the effective ones, appeared to be identical all the time as shown in panel (c) of Figs.\,\ref{fig:TT}A,B,C. Here, the system noise was controlled by tuning tolerance values only. Otherwise, there was no significant difference of implementation of the protocol in handling noises. Each unit cell consisted of eighteen particles (i.e., eight effective coordinates) and the space group appeared to be $Fd\bar{3}m$ confirming the cubic diamond unit cell. This suggested, the algorithm performed well towards the detection of unit cell from computer simulation data despite the presence of substantial noise. The somewhat distorted nature of the unit cell in the lowest density crystal was consistent with the same crystal structure and was the faithful representation of reality captured by the simulation. The fact that our method could withstand the several imperfections in the system just by modulating two parameters in a predictable manner, provided strong evidence in favor of correctness, as well as robustness of our algorithmic approach.

\begin{figure*}
	\centering 
	\includegraphics[scale=0.21]{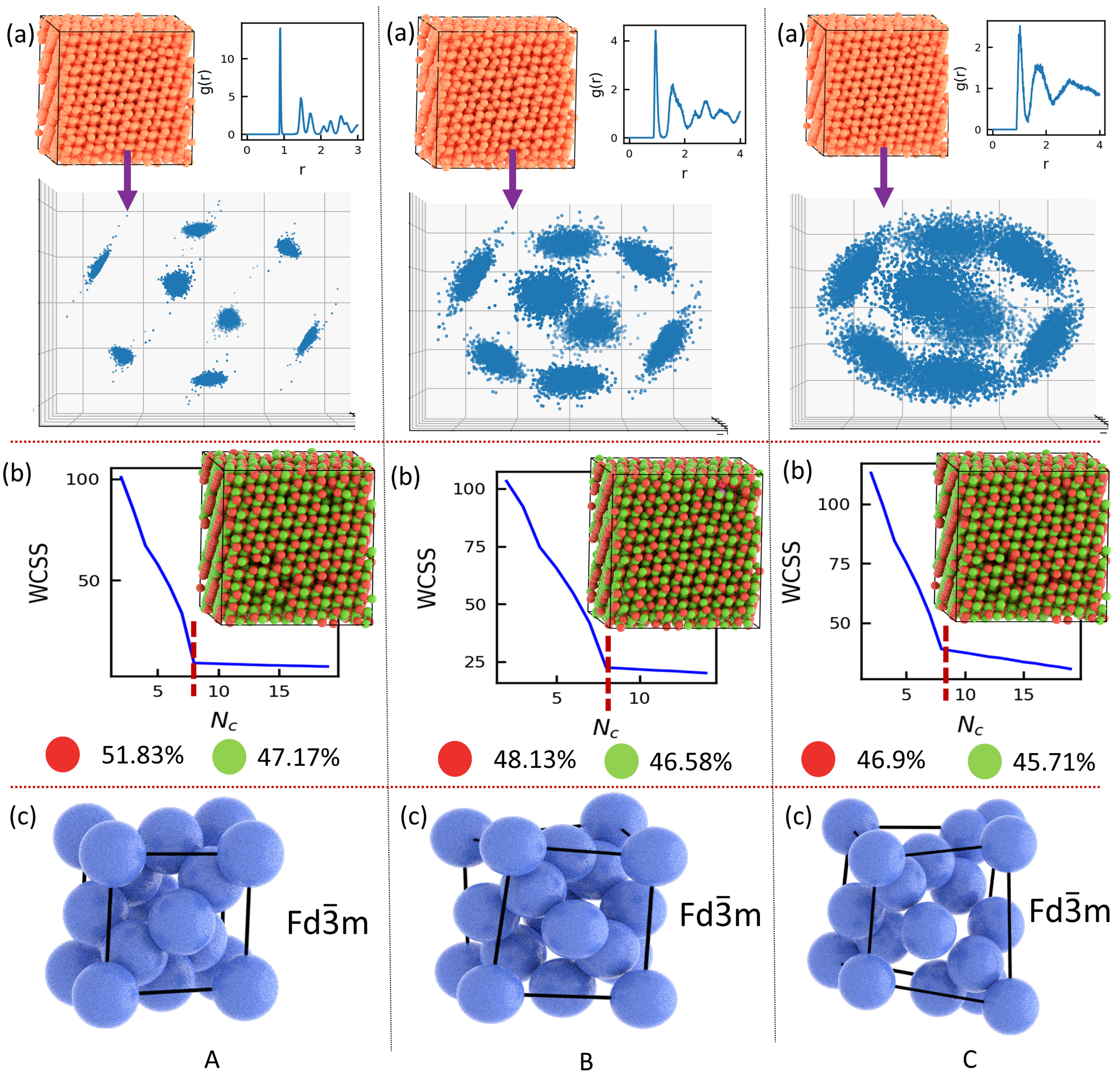}
	\caption{\textbf{Unit cell detection of simulated cubic diamond structures are shown for three different noise levels as depicted by the systems at three packing fractions:} \textbf{(A)} lower noise level ($\phi \sim$ 0.89), \textbf{(B)} moderate noise level ($\phi \sim$ 0.7) and \textbf{(C)} higher noise level ($\phi \sim$ 0.6). The configurations are shown in panel (a) along with the RDF analyses to calculate the cutoff distances followed by measuring the positional distributions of local neighbors. The number of clusters appear to be eight as confirmed by \textit{K-Means} clustering (shown in the top row of panel (b)). The configurations with two types of environments are shown with distinct colors in the insets of panel (b). The majority of particles with each kind of environment are mentioned in text indicating the existence of two types of environments in the system. Following the algorithm, cubic diamond unit cells are detected and shown in panel (c) with the space group as mentioned. It indicates the robustness of the algorithm despite adequate level of noises.}
	\label{fig:TT}
\end{figure*}

\section{Discussion}\label{sec:Discussion}
We report a direct approach to detect all possible unit cells and the space group of a crystallite formed in a typical computer simulation, based on the real-space analysis of particle coordinates. The method is based on a set heuristics derived from several well known and well utilized facts and careful observation of certain other features viewed through a different lens. The idea of BOO diagram was used extensively and furthermore, for a simple Bravais system, the uniqueness of the coordination polyhedron was utilized to deduce the crystal class. The next step exploited the observations that lattice vector directions for all known Bravais lattices had specific relationships with certain geometrical attributes of the coordination polyhedron. This gave rise to a limited number of triplets of lattice vectors directions to search for, which led to the eventual identification of the unit cell parameters and respective set of coordinates of effective particles inside each of the possible unit cells. The combinations that were crystallographically viable were put forward as the final output of the algorithm. The entire procedure relied on the concept of local environments, and standard computational techniques e.g.\,clustering, search and determination of convex hull, given that all unit cells are convex bodies. The whole approach worked because of judicious application of different elements in a specific sequence and ultimately choosing the answers that were consistent with crystallography.

This algorithm was, however, strictly valid for simple Bravais lattices and could not be used to determine the unit cell for general systems with complex bases. This led us to the second important component of the overall method, namely the ``environment separation''. Without this step, the algorithm for detecting unit cells from coordination polyhedron could not be used. The main reason behind this was the lack of empirical relationship between the directions of lattice vectors and the coordination polyhedron, which was only valid for a Bravais system. The environment separation iteratively transformed a general system to a Bravais lattice, while focusing on only one type of particles. This fact, i.e.\,a crystal can be thought of as a collection of Bravais systems, is not new knowledge, but, it has not been used in the context of crystal structure detection. The transformed Bravis system could be analyzed to detect the crystal class and the envelope of the intended unit cell. Subsequently, constructing convex hulls involving the original coordinates, followed by determination of effective particle coordinates, the final set of unit cells were produced. From this point the determination of the space group was a trivial task, and was accomplished by the \textsl{Spglib} package \cite{spglibv1}. In summary, the success of the entire procedure depended on careful observation and efficient utilization of known crystallographic facts along with standard computational techniques. 

The detection of broken rotational symmetry in crystals was first formulated by Steinhardt \textit{et al.}\,\,in terms of bond orientational order \cite{Steinhardt1983}, which was originally proposed to distinguish local non-liquid like order in dense systems. Extension of this idea, with later modifications \cite{Lechner2008, Reinhard2012}, resulted in methods for detecting crystal structures. But they were unable to detect anything beyond simple crystals with FCC, BCC and HCP structures, let alone really complex crystals with multiparticle bases, e.g.\,$\beta$-Mn. The inability of the Steinhardt order parameters to detect complex crystals is not entirely surprising. They essentially rely on different mathematical representation of the BOO, i.e.\,only use the basic information from the coordination polyhedron, and it is not enough for resolution of the full crystal symmetry. Our approach should not be viewed as a simple extension of approaches based on Steinhardt order parameters. It is also not directly related to  Ackland-Jones order parameter \cite{Ackland2006}, only similarity being the use of local environments, which, as mentioned before, is not enough, explaining its usual limitations.

Among all the reported methods, our approach is most closely related to the polyhedral template matching method \cite{Larsen2016}, which explicitly uses the information from coordination polyhedron. Despite being quite successful for several structure type, the PTM seems to perform poorly in crystals with lower symmetry, e.g.\, orthorhombic, monoclinic, and triclinic. Our method is fundamentally different from recently reported machine learning based approaches \cite{Ziletti2018, Boattini2019, Leitherer2021}. All these methods are dependent on the quantity and quality of training data, among other things, which is not the case for the current approach, which is based on heuristics, but deals with each set of coordinates in a direct manner without relying on any additional information.

As identifying proper symmetry operations of an unknown crystal structure was not possible before the proper detection of unit cell, our approach did not require to evaluate the Wyckoff positions \cite{Kittel2005, Massa2004} of the input structure. Though the Wyckoff positions are one of the most common attributes used to identify the crystal structures, our protocol did not need this, as our target was to identify the entire unit cell directly from the particle coordinates.

The method is quite robust in terms of its ability to handle statistical noise typically found in particle simulations, as long as proper distance and angle tolerances are used. These values need to be manually adjusted, but in our experience, this did not appear to be a serious issue. If the method fails to provide a meaningful answer, despite scanning a range of reasonable values of the said cutoffs, the crystallinity of the input coordinates should be questioned. The method is indented for single crystalline samples, but variations around similar ideas could be developed to deal with polycrystallinity. It is also specifically designed for analysis of bulk crystals formed in computer simulations with periodic boundary conditions, and not immediately suitable for nucleation studies from dense fluid phases. If the crystal is not of sufficient dimension, the local neighbor analysis based on coordination polyhedron will be extremely noisy and subsequent search protocols will fail. However, several intuitive ideas, as well as algorithmic details used here could provide new frameworks for developing order parameters suitable for nucleation studies and such endeavors will be undertaken in near future. 

\section{Conclusion}\label{sec:Conclusion}
We presented an algorithm to detect the complete unit cell information of all possible crystallogrphically viable choices from direct analysis of real space coordinate data obtained from typical computer simulations. The method was capable of handling noise by employing appropriate cutoffs and could detect complex crystals with multiparticle bases. The key concepts behind the heuristic algorithm were symmetry of the coordination polyhedron, its relationship with lattice vector directions for simple Bravais lattices and the fact that any complex crystal could be decomposed into a collection of simple Bravais lattices. These ideas were judiciously exploited by implementing standard computational techniques, e.g.\,\,search, clustering and convex hull constructions. The entire procedure can be carried out with minimal human intervention by executing a collection of Python scripts and requires negligible computational resource for typical system sizes used in contemporary simulation studies. Therefore, the method is suitable for algorithmic analysis of large data sets. Extensions around these ideas could be pursued to handle polycrystalline samples and devise new order parameters for nucleation studies. Our work provides an efficient general solution to the longstanding problem of crystal structure detection in particle based simulations, and is likely to be beneficial to researchers working in broad areas of condensed matter physics and computational materials science.

\begin{acknowledgments}
We acknowledge financial support from DST-INSPIRE Fellowship (IVR No.\,201800024677) provided to SK. AD thanks DST-SERB Ramanujan Fellowship (SB/S2/RJN-129/2016) and IACS start-up grant. KC thanks IACS for financial support. Computational resources were provided by IACS HPC cluster and partial use of equipments procured under SERB CORE Grant No.\,CRG/2019/006418.
\end{acknowledgments}

\section*{AUTHOR DECLARATIONS}
\subsection*{Conflict of Interest}
The authors have no conflicts to disclose.

\subsection*{Author Contributions}
SK and AD designed the research. SK and KC executed the idea computationally and implemented on multiple systems. SK and AD wrote the paper.

\section*{DATA AVAILABILITY}
The data that support the computational approach of this method are available within the article.


\newpage

\begin{center}
	\LARGE Supplemental Material : Algorithmic detection of the crystal structures from computer simulation data \\
	
	\large Sumitava Kundu, Kaustav Chakraborty, Avisek Das
\end{center}

\newpage

\section{Conditions to detect the crystal class using the vectors obtained from the polyhedron}
A set of non-unit vectors were obtained joining the center (0, 0, 0) and vertices, face mid-points or edge mid-points of the constructed polyhedron. Each triplet of the  $\Comb{g}{3}$ combinations was chosen as dummy basis vectors designated by $\vec{a}_d$, $\vec{b}_d$, $\vec{c}_d$ as discussed in the main text. The crystal class of the Bravais lattice was checked using each triplet followed by a categorization of those according to the agreement with crystal class within the tolerances. The conditions of satisfying seven crystal classes based on the lattice parameters $|\vec{a}_d|$, $|\vec{b}_d|$, $|\vec{c}_d|$, $\alpha$, $\beta$, $\gamma$ are given below.

\begin{enumerate}
	\item Cubic $\rightarrow$ $||\vec{a}_d| - |\vec{b}_d|| \leq \mathcal{X}_d$, $||\vec{a}_d| - |\vec{c}_d|| \leq \mathcal{X}_d$, $||\vec{b}_d| - |\vec{c}_d|| \leq \mathcal{X}_d$, $|\alpha - \beta| \leq \mathcal{X}_a$, $|\gamma - \beta| \leq \mathcal{X}_a$, $|\alpha - \gamma| \leq \mathcal{X}_a$, $(90^\circ - \mathcal{X}_a) \leq \alpha \leq (90^\circ + \mathcal{X}_a)$, $(90^\circ - \mathcal{X}_a) \leq \beta \leq (90^\circ + \mathcal{X}_a)$, $(90^\circ - \mathcal{X}_a) \leq \gamma \leq (90^\circ + \mathcal{X}_a)$
	\item Hexagonal $\rightarrow$ (i) $||\vec{a}_d| - |\vec{b}_d|| \leq \mathcal{X}_d$, $||\vec{a}_d| - |\vec{c}_d|| > \mathcal{X}_d$, $||\vec{b}_d| - |\vec{c}_d|| > \mathcal{X}_d$, $(90^\circ - \mathcal{X}_a) \leq \alpha \leq (90^\circ + \mathcal{X}_a)$, $(90^\circ - \mathcal{X}_a) \leq \beta \leq (90^\circ + \mathcal{X}_a)$, $(120^\circ - \mathcal{X}_a) \leq \gamma \leq (120^\circ + \mathcal{X}_a)$ \\
	
	(ii) $||\vec{a}_d| - |\vec{c}_d|| \leq \mathcal{X}_d$, $||\vec{a}_d| - |\vec{b}_d|| > \mathcal{X}_d$, $||\vec{b}_d| - |\vec{c}_d|| > \mathcal{X}_d$, $(90^\circ - \mathcal{X}_a) \leq \alpha \leq (90^\circ + \mathcal{X}_a)$, $(120^\circ - \mathcal{X}_a) \leq \beta \leq (120^\circ + \mathcal{X}_a)$, $(90^\circ - \mathcal{X}_a) \leq \gamma \leq (90^\circ + \mathcal{X}_a)$ \\
	
	(iii) $||\vec{b}_d| - |\vec{c}_d|| \leq \mathcal{X}_d$, $||\vec{a}_d| - |\vec{b}_d|| > \mathcal{X}_d$, $||\vec{a}_d| - |\vec{c}_d|| > \mathcal{X}_d$, $(120^\circ - \mathcal{X}_a) \leq \alpha \leq (120^\circ + \mathcal{X}_a)$, $(90^\circ - \mathcal{X}_a) \leq \beta \leq (90^\circ + \mathcal{X}_a)$, $(90^\circ - \mathcal{X}_a) \leq \gamma \leq (90^\circ + \mathcal{X}_a)$ 
	\item Tetragonal $\rightarrow$ $\rightarrow$ (i) $||\vec{a}_d| - |\vec{b}_d|| \leq \mathcal{X}_d$, $||\vec{a}_d| - |\vec{c}_d|| > \mathcal{X}_d$, $||\vec{b}_d| - |\vec{c}_d|| > \mathcal{X}_d$, $|\alpha - \beta| \leq \mathcal{X}_a$, $|\gamma - \beta| \leq \mathcal{X}_a$, $|\alpha - \gamma| \leq \mathcal{X}_a$, $(90^\circ - \mathcal{X}_a) \leq \alpha \leq (90^\circ + \mathcal{X}_a)$, $(90^\circ - \mathcal{X}_a) \leq \beta \leq (90^\circ + \mathcal{X}_a)$, $(90^\circ - \mathcal{X}_a) \leq \gamma \leq (90^\circ + \mathcal{X}_a)$ \\
	
	(ii) $||\vec{a}_d| - |\vec{c}_d|| \leq \mathcal{X}_d$, $||\vec{a}_d| - |\vec{b}_d|| > \mathcal{X}_d$, $||\vec{b}_d| - |\vec{c}_d|| > \mathcal{X}_d$, $|\alpha - \beta| \leq \mathcal{X}_a$, $|\gamma - \beta| \leq \mathcal{X}_a$, $|\alpha - \gamma| \leq \mathcal{X}_a$, $(90^\circ - \mathcal{X}_a) \leq \alpha \leq (90^\circ + \mathcal{X}_a)$, $(90^\circ - \mathcal{X}_a) \leq \beta \leq (90^\circ + \mathcal{X}_a)$, $(90^\circ - \mathcal{X}_a) \leq \gamma \leq (90^\circ + \mathcal{X}_a)$ \\
	
	(iii) $||\vec{b}_d| - |\vec{c}_d|| \leq \mathcal{X}_d$, $||\vec{a}_d| - |\vec{b}_d|| > \mathcal{X}_d$, $||\vec{a}_d| - |\vec{c}_d|| > \mathcal{X}_d$, $|\alpha - \beta| \leq \mathcal{X}_a$, $|\gamma - \beta| \leq \mathcal{X}_a$, $|\alpha - \gamma| \leq \mathcal{X}_a$, $(90^\circ - \mathcal{X}_a) \leq \alpha \leq (90^\circ + \mathcal{X}_a)$, $(90^\circ - \mathcal{X}_a) \leq \beta \leq (90^\circ + \mathcal{X}_a)$, $(90^\circ - \mathcal{X}_a) \leq \gamma \leq (90^\circ + \mathcal{X}_a)$
	\item Trigonal $\rightarrow$ $||\vec{a}_d| - |\vec{b}_d|| \leq \mathcal{X}_d$, $||\vec{a}_d| - |\vec{c}_d|| \leq \mathcal{X}_d$, $||\vec{b}_d| - |\vec{c}_d|| \leq \mathcal{X}_d$, $|\alpha - \beta| \leq \mathcal{X}_a$, $|\gamma - \beta| \leq \mathcal{X}_a$, $|\alpha - \gamma| \leq \mathcal{X}_a$, $\alpha \leq (90^\circ - \mathcal{X}_a)$ or $\alpha > (90^\circ + \mathcal{X}_a)$, $\beta \leq (90^\circ - \mathcal{X}_a)$ or $\beta > (90^\circ + \mathcal{X}_a)$, $\gamma \leq (90^\circ - \mathcal{X}_a)$ or $\gamma > (90^\circ + \mathcal{X}_a)$
	\item Orthorhombic $\rightarrow$ $||\vec{a}_d| - |\vec{b}_d|| > \mathcal{X}_d$, $||\vec{a}_d| - |\vec{c}_d|| > \mathcal{X}_d$, $||\vec{b}_d| - |\vec{c}_d|| > \mathcal{X}_d$, $\rightarrow$  $|\alpha - \beta| \leq \mathcal{X}_a$, $|\gamma - \beta| \leq \mathcal{X}_a$, $|\alpha - \gamma| \leq \mathcal{X}_a$, $(90^\circ - \mathcal{X}_a) \leq \alpha \leq (90^\circ + \mathcal{X}_a)$, $(90^\circ - \mathcal{X}_a) \leq \beta \leq (90^\circ + \mathcal{X}_a)$, $(90^\circ - \mathcal{X}_a) \leq \gamma \leq (90^\circ + \mathcal{X}_a)$
	\item Monoclinic $\rightarrow$ (i)  $||\vec{a}_d| - |\vec{b}_d|| > \mathcal{X}_d$, $||\vec{a}_d| - |\vec{c}_d|| > \mathcal{X}_d$, $||\vec{b}_d| - |\vec{c}_d|| > \mathcal{X}_d$, $|\alpha - \beta| \leq \mathcal{X}_a$, $|\gamma - \beta| \geq \mathcal{X}_a$, $|\alpha - \gamma| \geq \mathcal{X}_a$, $(90^\circ - \mathcal{X}_a) \leq \alpha \leq (90^\circ + \mathcal{X}_a)$, $(90^\circ - \mathcal{X}_a) \leq \beta \leq (90^\circ + \mathcal{X}_a)$, $(90^\circ + \mathcal{X}_a) \leq \gamma$\\
	
	(ii) $||\vec{a}_d| - |\vec{b}_d|| > \mathcal{X}_d$, $||\vec{a}_d| - |\vec{c}_d|| > \mathcal{X}_d$, $||\vec{b}_d| - |\vec{c}_d|| > \mathcal{X}_d$, $|\alpha - \gamma| \leq \mathcal{X}_a$, $|\gamma - \beta| \geq \mathcal{X}_a$, $|\alpha - \beta| \geq \mathcal{X}_a$, $(90^\circ - \mathcal{X}_a) \leq \alpha \leq (90^\circ + \mathcal{X}_a)$, $(90^\circ + \mathcal{X}_a) \leq \beta$, $(90^\circ - \mathcal{X}_a) \leq \gamma \leq (90^\circ + \mathcal{X}_a)$\\
	
	(iii) $||\vec{a}_d| - |\vec{b}_d|| > \mathcal{X}_d$, $||\vec{a}_d| - |\vec{c}_d|| > \mathcal{X}_d$, $||\vec{b}_d| - |\vec{c}_d|| > \mathcal{X}_d$, $|\beta - \gamma| \leq \mathcal{X}_a$, $|\gamma - \alpha| \geq \mathcal{X}_a$, $|\alpha - \beta| \geq \mathcal{X}_a$, $(90^\circ + \mathcal{X}_a) \leq \alpha$, $(90^\circ - \mathcal{X}_a) \leq \beta \leq (90^\circ + \mathcal{X}_a)$, $(90^\circ - \mathcal{X}_a) \leq \gamma \leq (90^\circ + \mathcal{X}_a)$
	\item Triclinic $\rightarrow$ $||\vec{a}_d| - |\vec{b}_d|| > \mathcal{X}_d$, $||\vec{a}_d| - |\vec{c}_d|| > \mathcal{X}_d$, $||\vec{b}_d| - |\vec{c}_d|| > \mathcal{X}_d$, $|\alpha - \beta| > \mathcal{X}_a$, $|\alpha - \gamma| > \mathcal{X}_a$, $|\beta - \gamma| > \mathcal{X}_a$
	
\end{enumerate}

\newpage
\section{Algorithm to detect corner particles of the parallelepiped for the Bravais lattice}
Considering the directions of probable basis vectors, $\hat{a}$, $\hat{b}$, $\hat{c}$, it was possible to search for a particle located at the nearest distance along each of the three directions with respect to a reference particle in the system. This protocol is discussed in the paragraph ``a'' of Section III C(1) in the main text. The algorithm to accomplish all eight corner particles of the parallelepiped is discussed below.

\vspace*{0.5cm}
\hrule
\begin{algorithmic}
	\Require $\hat{a}, \hat{b}, \hat{c} \text{ and coordinates of all particles in Bravais lattice}$
	\State $P_{arr} = []$
	\State $\mathcal{L} = [\hat{a}, \hat{b}, \hat{c}]$
	\State $\text{Choose a random particle, \textit{c}}$
	\For{$l$ \textit{in} $\mathcal{L}$}
	\State $\mathcal{Q} = []$
	\For{$i$ $\rightarrow$ (1, $N$+1)}
	\State $\vec{v}$ = $\vec{r}_i - \vec{r}_c$
	\State $\hat{v} = \vec{v}/|\vec{v}|$
	\State $\theta = \cos^{-1}(\hat{v} \cdot l)$
	\If{$\theta \leq \mathcal{X}_a$}
	\State $\mathcal{Q}\text{.insert(}i\text{)}$
	\EndIf
	\EndFor
	\State $\text{dist }= []$
	\For{$k$ \textit{in} $\mathcal{Q}$}
	\State $d = \dfrac{\vec{r}_k - \vec{r}_c}{|\vec{r}_k - \vec{r}_c|}$
	\State $\text{dist.insert(}d\text{)}$ 
	\EndFor
	\State $d_{min} = \text{min(dist)}$
	\State $ind = \text{The index of } \mathcal{Q} \text{ for which } d_{min} = \text{ min(dist)}$
	\State $P_{arr}\text{.insert(}ind\text{)}$
	\EndFor
\end{algorithmic}
\hrule
\vspace*{0.3cm}

\newpage
\section{Algorithm to detect effective particle coordinates in the unit cell}
Upon identifying all particles in the unit cell, it was required to detect the coordinates of effective particles using the chosen basis vectors. The protocol mentioned in the paragraph ``c'' of Section III C(3) in the main text, was followed using the algorithm as discussed below.

\vspace*{0.5cm}
\hrule
\begin{algorithmic}
	\Require $\text{Coordinate of } N_{uc} \text{ particles }$
	\State $\mathcal{D} = []$
	\For{$i \rightarrow (1 \text{ , } N_{uc})$}
	\If {$i \text{\textit{ not in }} \mathcal{D}$}
	\For{$j \rightarrow ((i+1)\text{ , } N_{uc}+1)$}
	\If {$j \text{\textit{ not in }} \mathcal{D}$}
	\State $\vec{r}_{ij} = r_j - r_i$
	\State $\hat{r}_{ij} = \dfrac{\vec{r}_{ij}}{|\vec{r}_{ij}|}$
	\State $\Theta_1 = \cos^{-1}(\hat{r}_{ij} \cdot \hat{\mathbf{a}}), \Theta_2 = \cos^{-1}(\hat{r}_{ij} \cdot \hat{\mathbf{b}}), \Theta_3 = \cos^{-1}(\hat{r}_{ij} \cdot \hat{\mathbf{c}})$
	\If{$\Theta_1 \leq \mathcal{X}_a \text{ or } \Theta_2 \leq \mathcal{X}_a \text{ or } \Theta_3 \leq \mathcal{X}_a$}
	\If{$(|\vec{\mathbf{a}}|-\mathcal{X}_d) \leq |\vec{r}_{ij}| \leq (|\vec{\mathbf{a}}|+\mathcal{X}_d) \text{ or } (|\vec{\mathbf{b}}|-\mathcal{X}_d) \leq |\vec{r}_{ij}| \leq (|\vec{\mathbf{b}}|+\mathcal{X}_d) \text{ or } (|\vec{\mathbf{c}}|-\mathcal{X}_d) \leq |\vec{r}_{ij}| \leq (|\vec{\mathbf{c}}|+\mathcal{X}_d)$}
	\State $\mathcal{D}\text{.insert($j$)}$
	\EndIf
	\EndIf
	\EndIf
	\EndFor
	\EndIf
	\EndFor	
	\State $\mathcal{B} = []$
	\For{$i \rightarrow (1 \text{ , } N_{uc}+1)$}
	\If {$i \text{\textit{ not in }} \mathcal{D}$}
	\State $\mathcal{B}\text{.insert($i$)}$
	\EndIf
	\EndFor
\end{algorithmic}
\hrule 
\vspace*{0.3cm}

\newpage
\section{Algorithm to convert a non-Bravais lattice into a Bravais lattice}
A non-Bravais lattice was required to transform into a Bravais lattice upon the detection of original system to have multiple kinds of local environments within the cutoff $r_c$. This has been discussed elaborately in the Sections IV B,C of the main text. The required algorithm is presented indicating all the necessary steps.

\vspace*{0.5cm}
\hrule
\begin{algorithmic}
	\Require $N \text{ particle coordinates}$
	\State $\text{Choose $r_c$ from RDF}$
	\State $\text{BOO within $r_c$}$
	\State $\text{\textit{K-Means} clsutering of BOO}$
	\State $\text{Calculate } \underline{\mathbb{E}}_i \text{ for all particles}$
	\State $\text{Separate } N_{s} \text{ particles based on } \underline{\mathbb{E}}_i \text{, where } i \rightarrow 1\text{, } N+1$
	\While{$N_{s}$ $<$ $N$}
	\State $N \gets N_{s}$
	\State $\text{Updated system with } N \text{ particles}$
	\State $\text{Choose $r_c$ from RDF}$
	\State $\text{BOO within $r_c$}$
	\State $\text{\textit{K-Means} clsutering of BOO}$
	\State $\text{Calculate } \underline{\mathbb{E}}_i \text{ for all particles in the updated system}$
	\State $\text{Separate } N_{s} \text{ particles based on } \underline{\mathbb{E}}_i \text{, where } i \rightarrow 1\text{, } N+1$
	\EndWhile
\end{algorithmic}
\hrule
\vspace*{0.3cm}

\newpage
\section{Methods}\label{sec:methods}
We report the algorithmic approach to detect the unit cell of the crystal crystal from the particle coordinates only. All necessary steps of algorithm are discussed in the main text. Here, we discuss a few additional methods including the usages of multiple packages which were used to develop the algorithm. 

\subsection{Preparation of ideal crystal structures from the unit cell}\label{ideal_crys}
A Crystallographic Information File (CIF) file was chosen from the database \cite{Downs2003, Grazulis2009, Grazulis2015, Merkys2016, Merkys2023, Quiros2018, Vaitkus2023} corresponding to a particular crystal structure with provided lattice parameters and space group. The unit cell was visualized and replicated 7 times ($n$=7) along the directions of the basis vectors in three dimensions using the \textsl{Ovito} package \cite{Stukowski2010}. In this way, the ideal crystal structure was achieved with the exact lattice parameters and space group. All the information other than the particle coordinates were lost intentionally and used as the only inputs of the algorithm. Our aim was to extract the unit cell from the crystal structure and compare all details with the unit cell which was replicated to prepare the crystal initially. In this way, ideal crystal structures of Barium (\ch{Ba}), Strontium Oxalate (\ch{C3O6Sr}), Potassium Sulfate (\ch{K2SO4}) and Iron magnesium oxide (\ch{Fe2MgO4}) were prepared and respective data are reported here along with two examples discussed in the main text.

\subsection{Simulation details for producing the cubic diamond crystals}\label{sim_details}
A dilute system (packing fraction $\phi$ $\sim$ $0.10$) of regular Truncated Tetrahedron shape was slowly compressed by constant volume Monte Carlo (MC) simulations up to a packing fraction of $\sim 60\%$ under the influence of hard-core interaction using HOOMD-Blue simulation toolkit \cite{Anderson2016a}. During each compression cycle, first all overlaps were removed, followed by a short MC run with the current volume. At the end of the compression stage the systems were simulated under the constant volume condition until spontaneous crystallization was observed. The crystalline solid was further equilibrated under constant pressure calculation allowing full anisotropic fluctuations of the simulation box. The reduced pressure $p^{\ast} = \beta p v_0$, where $\beta = (k_{B} T)^{-1}$ and $v_0$ is particles volume, was calculated from the constant volume crystallization simulation by the scaled distribution function method implemented in HOOMD. The particle volume $v_0$ was set to 1.0 to compare the system density with packing fraction. This value of pressure was taken as a base value and the system was further compressed by slowly ramping the external pressure, at each pressure value long equilibration run was carried out to ensure stable fluctuations of the box dimensions around the mean values. This process was continued to the a high value of external pressure beyond which no noticeable decrease in volume was observed. The densest packing ($\phi \sim 0.87$) of regular Truncated Tetrahedron shape under hard-core interaction exhibited cubic diamond crystal structure at very high value of $p^{\ast}$ as reported in the literature \cite{Damasceno2012c}. The entire process typically involved $\sim 20$ stages. The most compressed state obtained at the highest value of pressure ($p^{\ast}$=100) was taken as the starting point for a series of fresh set of constant pressure simulations where the pressure was reduced slowly, until the system completely melted into an isotropic liquid phase. The entire melting regime was simulated at same pressure values where the compression simulations were performed. The crystal structures produced at the densest state ($\phi \sim 0.87$) and other two packing fractions during melting ($\phi \sim$ 0.70, 0.60) were used to implement the algorithm and the data are reported in main text.

\subsection{Bond-orientational order within a cutoff distance}\label{boo}
Bond-orientational order (BOO) is a measure of local translational and orientational broken symmetry in a crystal. Considering a reference particle (for example, $i$-th particle), all the neighbors were identified within a certain distance $r_c$ (as discussed in the main text) followed by measuring the relative positional vectors from the reference particle to all the neighbors, $\vec{r}_{ij}$ = $\vec{r}_j$ - $\vec{r}_i$, where $i$ $\in$ $1, \ldots, N$ and $j \in 1, \ldots, N_{CN}$, $N$ and $N_{CN}$ being the total number of particles in the system and number of neighbors for each particle respectively. The BOO order corresponds to the translationally superposed state of all vectors $\vec{r}_{ij}$ which measures the local broken symmetry state in the crystal structure. We measured the BOO using \textsl{freud}-toolkit \cite{freud2020} by providing the value of $r_c$ which requires human intervention as mentioned in the main test.

\subsection{Estimation of number of clusters using \textsl{K-Means} clustering }\label{kmeans}
We estimated the number of clusters present in the BOO using an algorithmic approach which was necessary for determining the particle clouds in a noisy crystal. The measurement was performed using \textsl{K-Means} clustering method \cite{MacQueen1967} implemented in \textsl{Scikit-learn} \cite{scikit-learn}. It was required to provide an initial guess of minimum ($c_{min}$) and maximum ($c_{max}$) number of clusters which needed manual monitoring. Based on the right choices of two values $c_{min}$ and $c_{max}$, the ``elbow analysis'' allowed us to measure sum of the square distance, \textsl{WCSS} (Within-Cluster Sum of Square) between the particles in a cluster and the centroids in order to calculate the number of clusters indicated by the ``elbow'' of the plot.

\newpage
\section{Complete algorithmic steps applied for ideal $\beta$-Mn crystal}
The $\beta$-Mn crystal structure was used to demonstrate the algorithm in the main text. The corresponding data are shown Fig.\,\ref{fig:betamn} indicating each step to explain the approach step by step. This made it more realizable while visualizing all the steps in a single figure.

\begin{figure*}[!h]
	\centering 
	\includegraphics[scale=0.23]{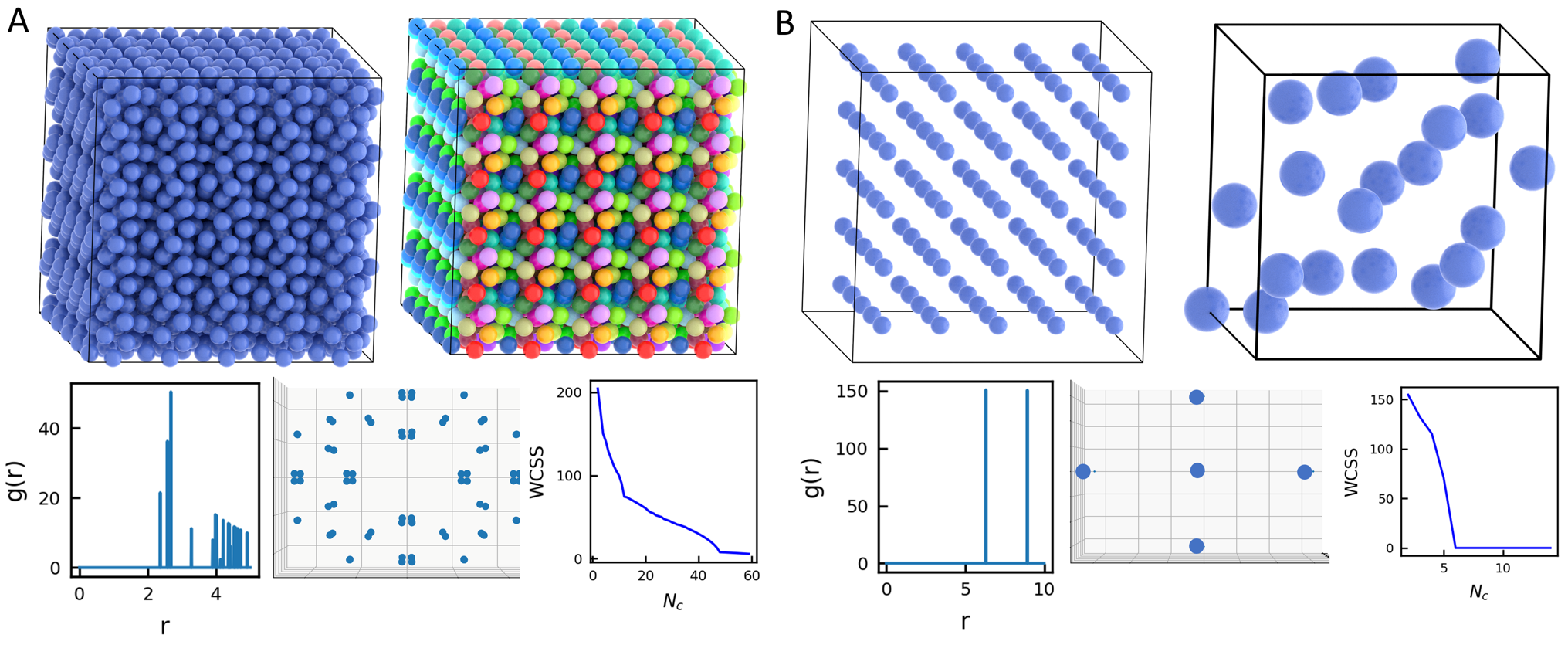}
	\caption{\textbf{All steps to detect the $\beta$-Mn unit cell are shown in detail.}  An ideal beta-Mn crystal structure shown in the top row of panel \textbf{(A)} (left figure) is considered followed by the determination of cutoff distance $r_c$ (=2.6) from radial distribution function, bond-orientational diagram and number of clusters ($N_c$=48) using \textit{K-Means} clustering method. The data are presented in the bottom row of panel \textbf{(A)}. As a part of the environment separation technique, the particles are colored based on the similar kind of environment and shown in the top row of panel \textbf{(A)} (right one). The particles sharing the similar kind of environment are separated out as shown in the top row of panel \textbf{(B)} (left one). The similar analyses displayed in the bottom row of \textbf{(B)} confirm the presence of six clusters within the new cutoff $r_c$=7.0 for the Simple Cubic Bravais lattice. The identified unit cell with 20 effective particle coordinates are shown in the top row of \textbf{(B)} (right one).}
	\label{fig:betamn}
\end{figure*}

\newpage
\section{Detection of unit cell of $\beta$-Mn crystal by varying the noise}
Gaussian white noise was added to each particle position independently in an ideal $\beta$-Mn crystal. The extent of noise was characterized by the standard deviation of the Gaussian distribution, $\sigma$, for which four values were used ($\sigma$ = 0.025, 0.05, 0.075, 0.1), which resulted in progressively noisy crystals (snapshots and BOO diagrams are presented in Figs.\, \ref{fig:betamn_sigma_0.025}, \ref{fig:betamn_sigma_0.05}, \ref{fig:betamn_sigma_0.075}, \ref{fig:betamn_sigma_0.1},along with relevant pictures of crucial steps as presented before). The results of the structure determination are summarized in Table \ref{table:noise_comp}, which confirms the success of the method in all foure scenarios. The percentage of particles with identical environment, a crucial quantity in our algorithm, did not appear to change significantly with the addtion of the noise. This was the main reason why all the steps works predictably, especially the selection of cutoffs and clustering. The only parameters that needed some manual adjustment were the distance and angular cutoffs ($\mathcal{X}_d$ and $\mathcal{X}_a$), and as shown in Table \ref{table:noise_comp}, the cutoff values were significant relaxed for successful determination of the structure. However, choosing them was relatively straightforward, despite the fourfold increase in noise.

\begin{figure*}[!h]
	\centering 
	\includegraphics[scale=0.23]{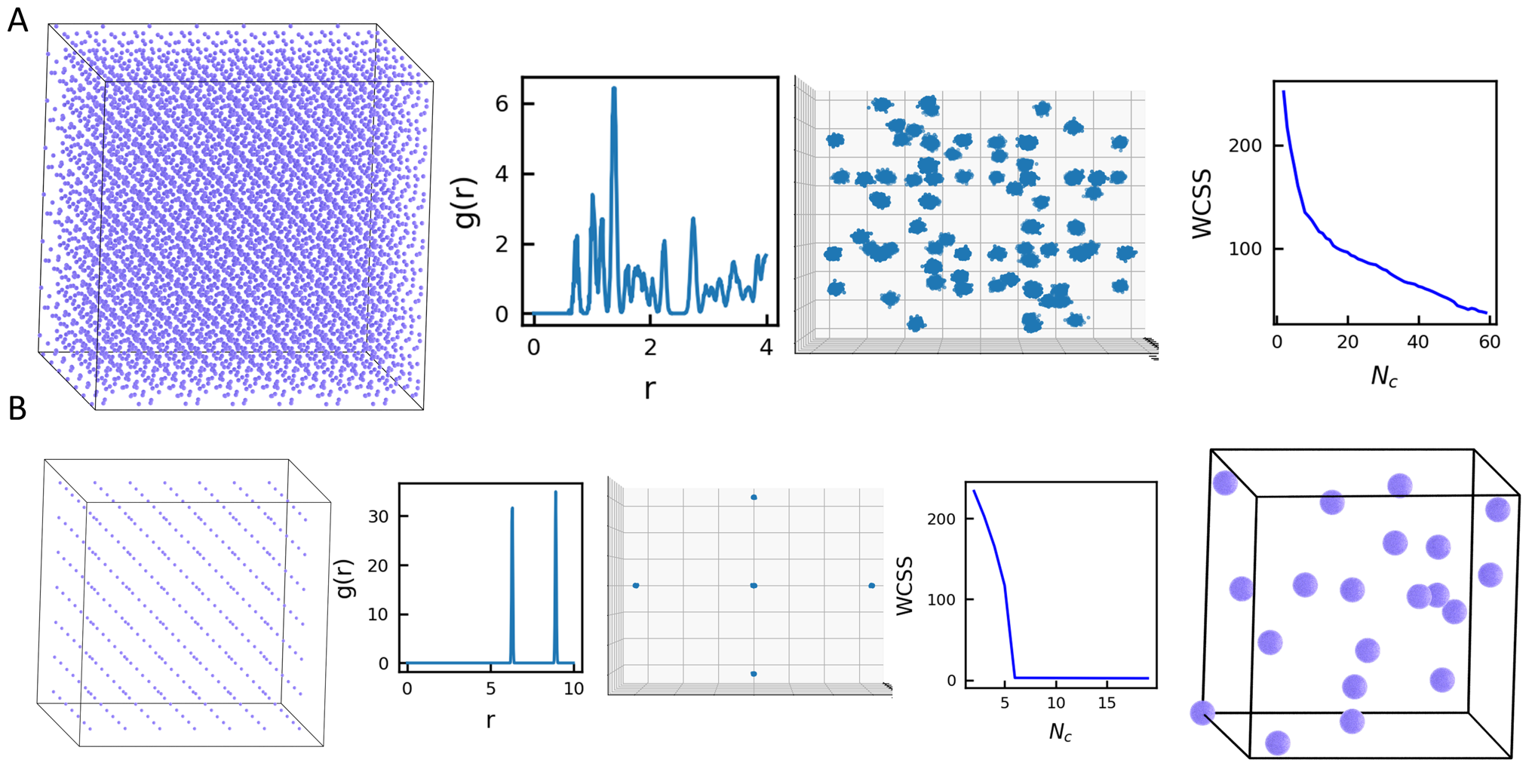}
	\caption{\textbf{Implementation of the algorithm are shown for $\beta$-Mn crystal at $\sigma$=0.025.} All the necessary steps are shown indicating the robustness of the algorithm to deal with the noise.}
	\label{fig:betamn_sigma_0.025}
\end{figure*}
\begin{figure*}
	\centering 
	\includegraphics[scale=0.23]{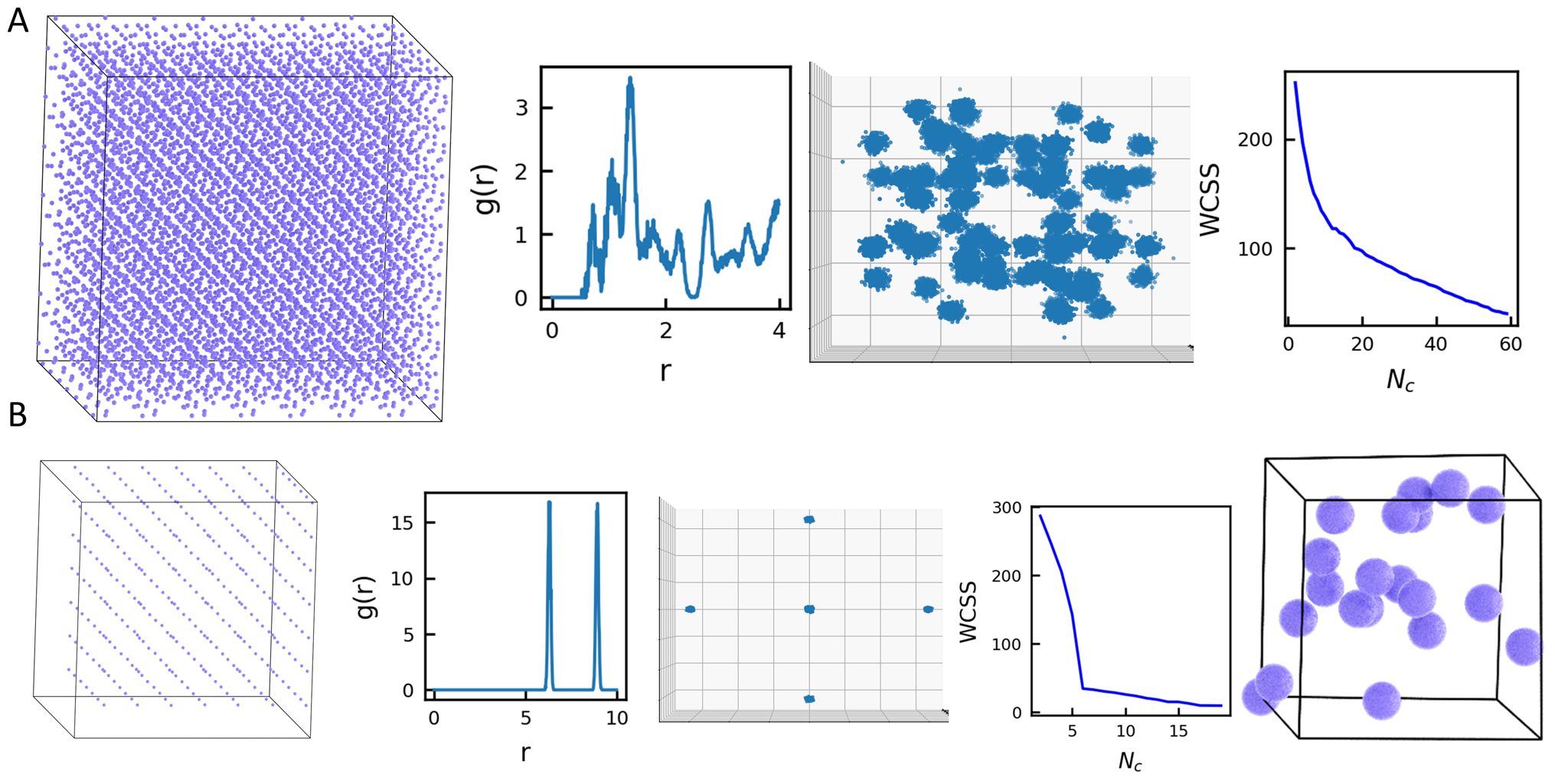}
	\caption{\textbf{Implementation of the algorithm are shown for $\beta$-Mn crystal at $\sigma$=0.05.} All the necessary steps are shown indicating the robustness of the algorithm to deal with the noise.}
	\label{fig:betamn_sigma_0.05}
\end{figure*}
\begin{figure*}
	\centering 
	\includegraphics[scale=0.23]{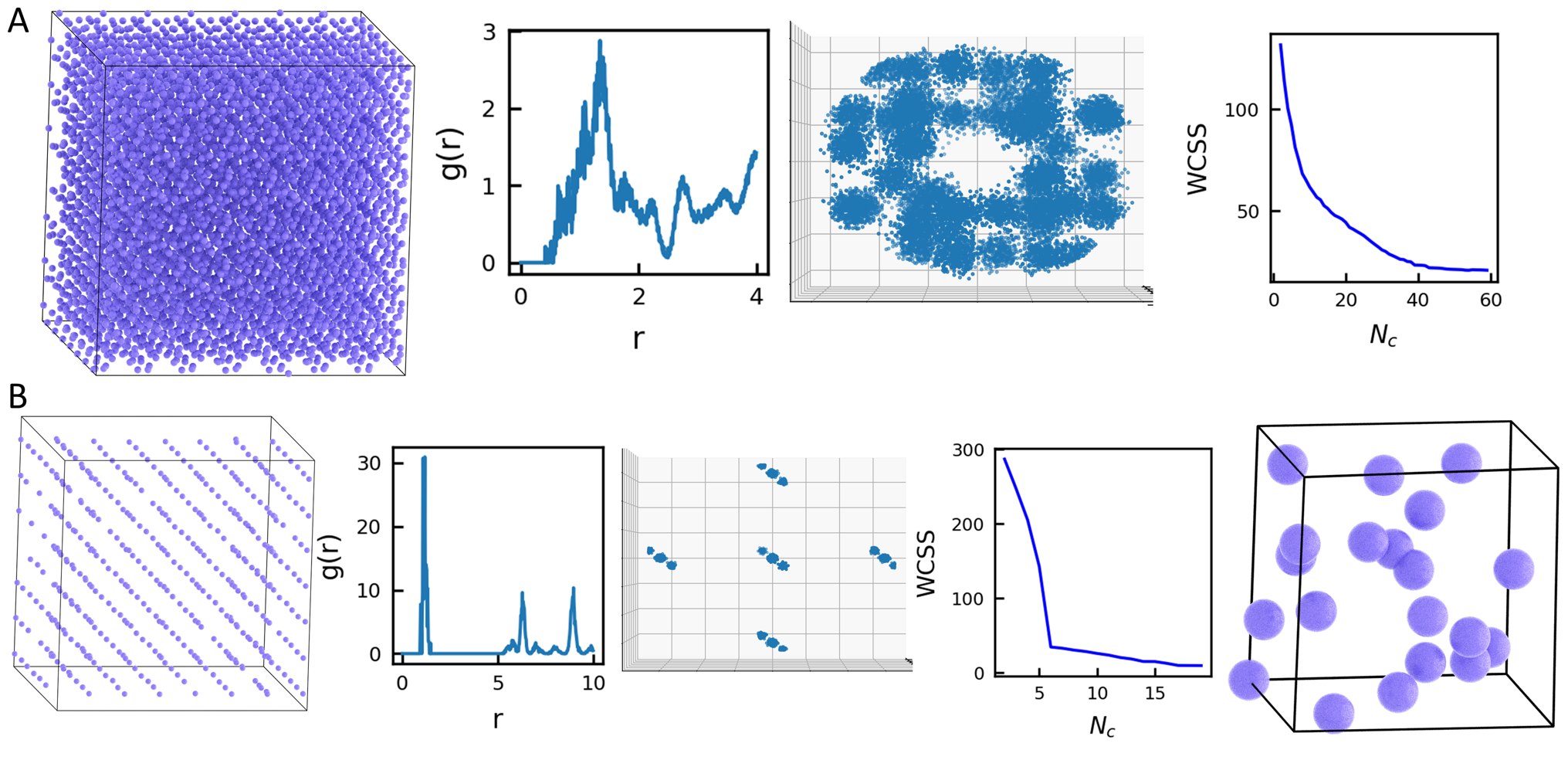}
	\caption{\textbf{Implementation of the algorithm are shown for $\beta$-Mn crystal at $\sigma$=0.075.} All the necessary steps are shown indicating the robustness of the algorithm to deal with the noise.}
	\label{fig:betamn_sigma_0.075}
\end{figure*}
\begin{figure*}
	\centering 
	\includegraphics[scale=0.23]{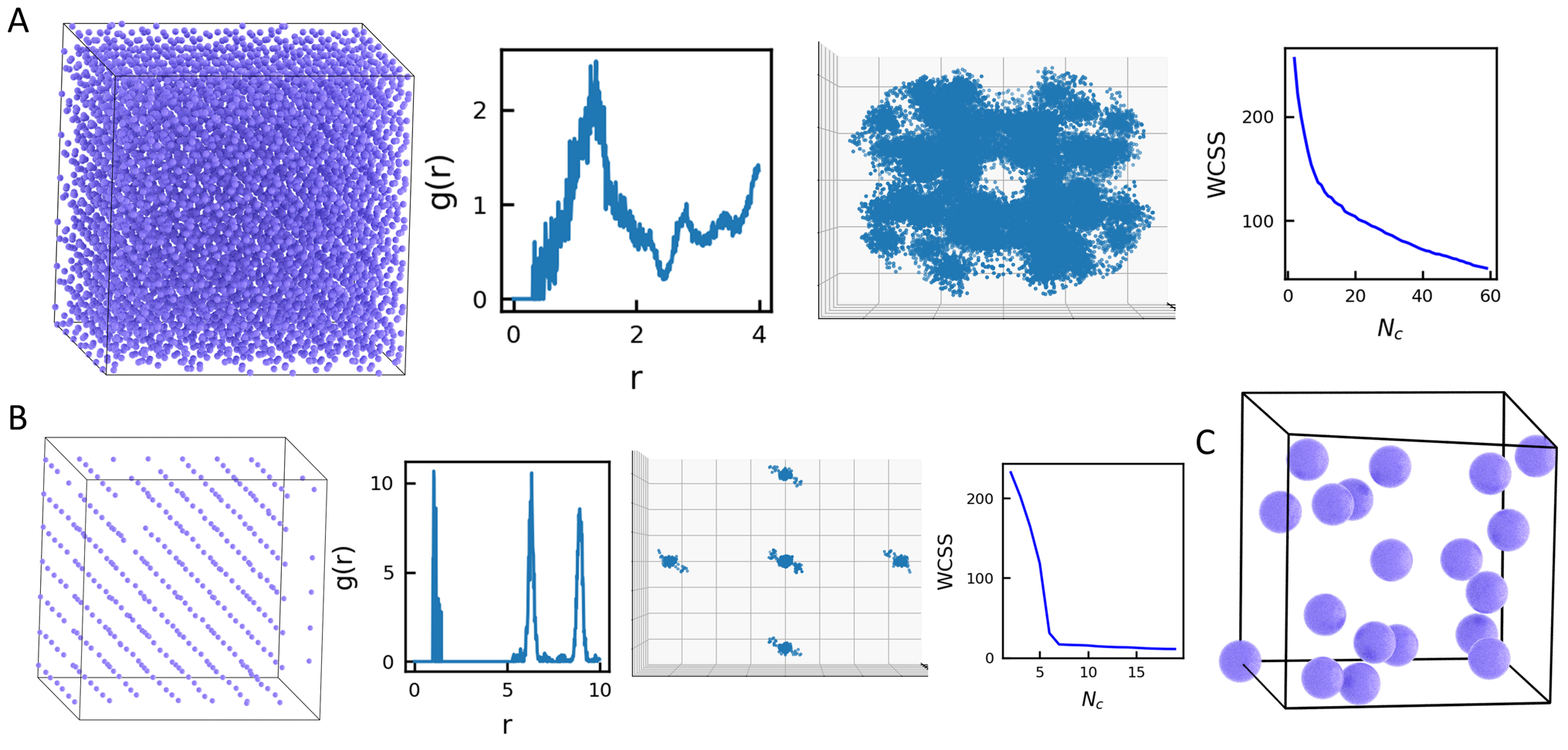}
	\caption{\textbf{Implementation of the algorithm are shown for $\beta$-Mn crystal at $\sigma$=0.1.} All the necessary steps are shown indicating the robustness of the algorithm to deal with the noise.}
	\label{fig:betamn_sigma_0.1}
\end{figure*}

\begin{table*}[t] 
	\centering
	{\renewcommand{\arraystretch}{1.0}%
		\begin{tabular}[t]{|P{2cm}|P{2cm}|P{2cm}|P{2cm}|P{3.5cm}|P{2cm}|}
			\hline
			Standard deviation of Gaussian noise ($\sigma$) & Particles with similar environment (\%) & Distance tolerance ($\mathcal{X}_d$) & Angle tolerance ($\mathcal{X}_a$) & Lattice parameters ($a$, $b$, $c$, $\alpha$, $\beta$, $\gamma$) & Space group \\
			\hline
			0.025 & 5 & 0.03 & 3$^\circ$ & 6.32, 6.31, 6.31, 89.89$^\circ$, 89.57$^\circ$, 89.83$^\circ$ & $P4_{1}32$ \\
			\hline
			0.05 & 5 & 0.05 & 5$^\circ$ & 6.32, 6.29, 6.3, 90.06$^\circ$, 90.46$^\circ$, 91.30$^\circ$ & $P4_{1}32$ \\
			\hline
			0.075 & 5.06 & 0.15 & 10$^\circ$ & 6.31, 6.28, 6.37, 87.80$^\circ$, 89.10$^\circ$, 92.77$^\circ$ & $P4_{1}32$ \\
			\hline
			0.1 & 4.98 & 0.2 & 10$^\circ$ & 6.21, 6.368, 6.39, 93.51$^\circ$, 91.41$^\circ$, 86.19$^\circ$ & $P4_{1}32$ \\
			\hline
		\end{tabular}
		\caption{The data upon the implementation of algorithm in the $\beta$-Mn crystal structure are shown at different noise levels. The standard deviation ($\sigma$) of the Gaussian noise is varied and corresponding tolerances are listed along with the detected lattice parameters and identical space group as the ideal system.}
		\label{table:noise_comp}
	}
\end{table*}

\newpage
\section{Implementation of the algorithm for other crystal structures}
We implemented the algorithm discussed in the main text in other two crystal structures to study the relevance of the method and report the results in the following two subsections. These two structures correspond to a body-centred cubic (BCC) crystal and Hexagonal-closed packed (HCP) structure.

\subsection{Test case 1 : Ideal monoclinic crystal structure of Strontium oxalate with space group $P_{1}2_{1}/C_{1}$} \label{sec:calcilum_oxalate_test}
An ideal crystal structure of strontium oxalate (\ch{C3O6Sr}) (space group - $P_{1}2_{1}/C_{1}$ ) was prepared synthetically by replicating the unit cell with lattice parameters, $a$ = 7.966, $b$ = 9.205, $c$ = 7.319, $\alpha$ = 90$^{\circ}$, $\beta$ = 102.104$^{\circ}$ and $\gamma$ = 90$^{\circ}$ \cite{Vanhoyland2001}. Fig.\,\ref{fig:C3O6Sr}A exhibits the exact configuration of the system with all the atoms in different colors and sizes where the corresponding unit cell is shown as inset. Following the algorithmic protocol, any particular types of atoms could be separated to accomplish the step. Here, we separated all \ch{Sr} atoms as presented in Fig.\,\ref{fig:C3O6Sr}B and considered this system to observe the positional distributions of neighbors following the recipe as designed in section III A of the main text. The value of $r_c$ ($\sim$ 5.0) was determined from the RDF analysis. In this case, the first minimum ($\sim$ 4.0) did not serve the purpose of a valid $r_c$ leading to the particles distributed only in two dimensions. We were allowed to choose any higher value of cutoff distance but that would slow down the convergence of algorithm as discussed earlier. The calculation of BOO and \textit{K-Means} clustering were carried out following the protocols described in the section III A of the main text as corresponding data are shown in panel Fig.\,\ref{fig:C3O6Sr}C. The existence of eight clusters was observed. Further analyses confirmed the system as a non-Bravais lattice with two kinds of environments for the particles; the configuration is shown in Fig.\,\ref{fig:C3O6Sr}D. In this configuration, the particles were portrayed in two colors indicating the similar ones to have identical environment. Thereafter, the particles with similar environment were separated following to the steps designed in section IV C of the main text as presented in Fig.\,\ref{fig:C3O6Sr}C. This system appeared to be a primitive Monoclinic Bravais lattice which was taken into account to analyze the directions of basis vectors. Similar analyses including the positional distributions of local neighbors within $r_c \sim 9.8$ were carried out followed by determining the number of clusters in BOO as shown in Fig.\,\ref{fig:C3O6Sr}F. The \textit{K-Means} clustering confirmed the appearance of eight clusters.  This transformed system was scrutinized further to check the Bravais or non-Bravais nature by performing similar analyses with the value of $r_c$ set at 6.0. The data are shown in the respective figures in the lower panel of Fig.\,\ref{fig:C3O6Sr}C. Subsequently, the steps outlined in the section III of the main text, were performed which led to the detection of a set consisting of eight unique choices of basis vectors. Among all relevant choices, one triplet matched with the considered one i.e., $\vec{a}$=[-2.448, 0, -7.157], $\vec{b}$=[-4.750, 0, 3.578] and $\vec{c}$=[0.0, -9.205, 0.0] and corresponding lattice parameters appeared to be 7.966, 9.205, 7.319, 90$^{\circ}$, 102.104$^{\circ}$, 90$^{\circ}$. Considering the coordinates of all atoms in the original system (snapshot shown in Fig.\,\ref{fig:C3O6Sr}A), the unit cells were identified with respective particle coordinates for all eight choices. The unit cell agreeing with the initially considered one, is shown in Fig.\,\ref{fig:C3O6Sr}G with 40 effective coordinates. The other choices also led to the existence of 40 effective coordinates in the cell with different lattice parameters and basis vectors. All the choices produced similar fractional coordinates and the space group appeared to be $P_{1}2_{1}/C_{1}$ each time within the tolerances, \textsl{symprec} = 0.01 and \textsl{angle\_tolerance} = 0.01$^\circ$, as set for ideal crystal. It suggested successful implementation of our algorithm to detect the unit cell of a complex crystalline structure with very number of symmetry elements.

\begin{figure*}[!h]
	\centering 
	\includegraphics[scale=0.23]{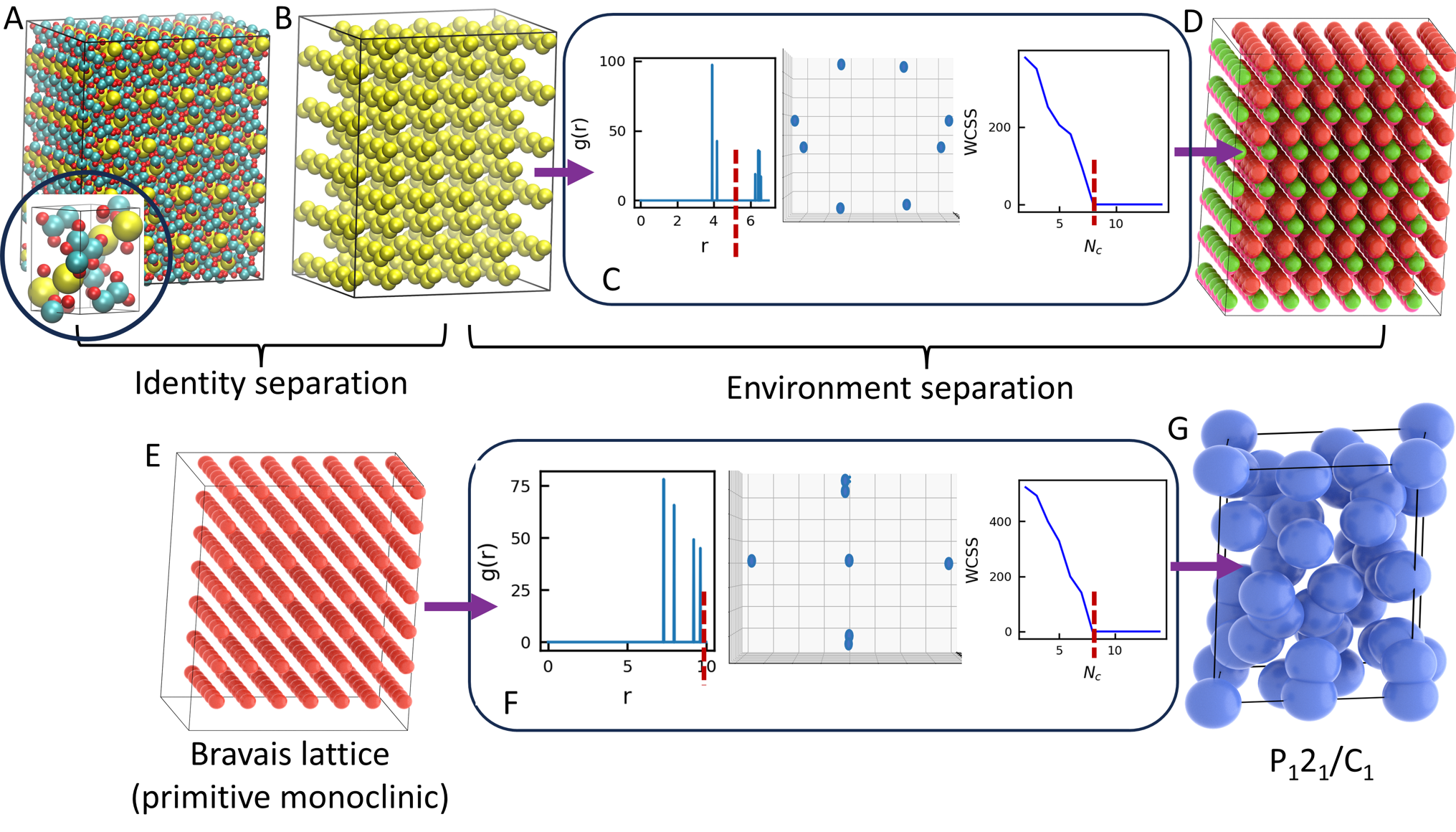}
	\caption{\textbf{Detection of Strontium Oxalate (\ch{C3O6Sr}) crystal structure are shown.} Panel \textbf{(A)} shows the configuration of ideal crystal with all atomic information and respective unit cell as inset. All \ch{Sr} atoms are separated and shown in panel \textbf{(B)}. The positional distribution of local neighbors within the cutoff $r_c \sim$ 5.0 (``red'' dotted line in RDF analyses) produces eight clusters in BOO as confirmed by \textit{K-Means} clustering. These analyses are presented in panel \textbf{(C)}. The existence of two types of environment assured the system as non-Bravais lattice and shown in panel \textbf{(D)}. The system turns into primitive Monoclinic Bravais lattice upon separation of the particles with similar environment as displayed in panel of \textbf{(E)}. The positional distributions are re-evaluated within the cutoff $r_c \sim$ 9.8 (``red'' dotted line in RDF analyses) confirming the existence of eight clusters as shown in panel \textbf{(F)}. The detected unit cell displayed in panel \textbf{(G)} confirms the space group of Strontium Oxalate crystal (written in text) with 40 effective coordinates indicating the versatility of approach in handling the crystal structure with lower number of symmetry elements.}
	\label{fig:C3O6Sr}
\end{figure*}

\subsection{Test case 2 : BCC crystal of Barium with space group $Im\bar{3}m$}
An ideal BCC crystal of Barium is shown in Fig.\,\ref{fig:Ba}A which was prepared using the unit cell with lattice parameters $a$=4.758, $b$=4.758, $c$=4.758, $\alpha$=90$^{\circ}$, $\beta$=90$^{\circ}$, $\gamma$=90$^{\circ}$ \cite{Kitano2001} (inset of (A)), followed by a calculation of RDF of the system, showing crystal-like nature (Fig.\,\ref{fig:Ba}B). The nearest neighbor distance ($r_c$) is calculated from the first minima of the RDF which turned out to be $\sim$ 4.5 (exact nearest neighbor distance was 4.12). Further, considering all pairwise distances, the positional distributions of the neighbors within $r_c$ is shown in (Fig.\,\ref{fig:Ba}C), formed eight particles (zero positional deviation for the ideal system) in three-dimensional space, as confirmed by the elbow analysis of \textit{K-Means} clustering algorithm as in Fig.\,\ref{fig:Ba}D. The environment separation was performed and it turned out that $\sim$ 100\% particles had similar kind of environment confirming the system as a Bravais lattice already. For this example, the transformation from a non-Bravais lattice into a Bravais lattice was not required. The coordinates of the centroid of the clusters (here the coordinate of the particles in the BOO) were considered as the vertices of a convex polyhedron. After the decomposition of the convex polyhedron, the vertices forming the polyhedron's faces and edges were evaluated followed by the calculations of determining the coordinates of the six face mid-points and twelve edge mid-points. Total number of non-unit vectors ($g$) calculated from the polyhedron was 26. Each triplet of $\Comb{26}{3}$ combinations was treated as a set of lattice vectors ($\vec{\mathbf{a}}_d$, $\vec{\mathbf{b}}_d$ and $\vec{\mathbf{c}}_d$) and the lattice parameters, $a$, $b$, $c$, $\alpha$, $\beta$ and $\gamma$ were determined. Computationally we checked each set of lattice parameter to know the crystal class satisfied by the those. Cubic crystal class was satisfied by few triplets as the highest order of symmetry i.e., 48. The directions of basis vectors appeared to be unique satisfying the cubic class and $\hat{\mathbf{a}}$ = [1, 0, 0], $\hat{\mathbf{b}}$ = [0, 1, 0] and $\hat{\mathbf{c}}$ = [0, 0, 1]. No other directions produced any valid Bravais lattices under the cubic class. Construction of the unit cell with the basis vectors followed the determination of actual lattice vectors, $\vec{\mathbf{a}}$ = [4.758, 0, 0], $\vec{\mathbf{b}}$ = [0, 4.758, 0] and $\vec{\mathbf{c}}$ = [0, 0, 4.758] with lattice parameters 4.758, 4.758, 4.758, 90$^{\circ}$, 90$^{\circ}$, 90$^{\circ}$. The BCC unit cell with eight particles at corners and one particle at the centre of the cubic box was obtained as shown in Fig.\,\ref{fig:Ba}E. The two effective particles were obtained and the space group detection technique was applied  with four particle basis and the space group turned out to be $Im\bar{3}m$ considering \textsl{symprec} = 0.0 and \textsl{angle\_tolerance} = 0$^\circ$.

\begin{figure*}
	\centering 
	\includegraphics[scale=0.8]{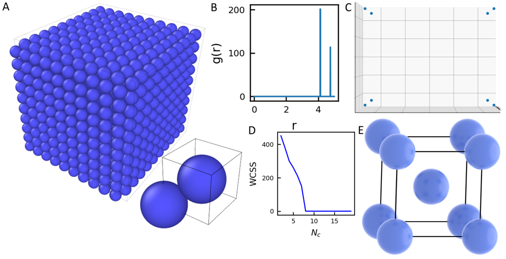}
	\caption{\textbf{Detection of unit cell for a BCC crystal is shown in multiple steps.} \textbf{(A)} An ideal BCC crystal of Barium (\ch{Ba}) is shown along with the unit cell in the inset. RDF and the distribution of the first nearest-neighbors are shown in \textbf{(B)} and \textbf{(C)} respectively with $r_c$ $\sim$ 4.5. There are eight clusters as confirmed by elbow analysis of the \textit{K-Means} clustering \textbf{(D)}. The BCC unit cell contains total nine particles as shown in \textbf{(E)}.}
	\label{fig:Ba}
\end{figure*}

\newpage
\subsection{Test case 3 : Hexagonal closed packed (HCP) structure of Potassium Sulfate with space group $Im\bar{3}m$}
Another example pertains to the detection of the unit cell and a space group of a synthetically prepared ideal Hexagonal-closed packed (HCP) structure of Potassium Sulfate (\ch{K2SO4}) with lattice parameters 5.947, 5.947, 8.375, $90^{\circ}$, $90^{\circ}$, $120^{\circ}$ and space group $P6_{3}/mmc$, having 22 effective particles in the unit cell \cite{Miyake1980}. The coordinates of all atoms irrespective of atom types were considered as the inputs of the algorithm. The data corresponding to all required steps are illustrated in Fig.\,\ref{fig:K2SO4}. As the first step of the layout, the crystal system was checked which appeared to be non-Bravais in nature as revealed by the analyses performed to detect the positional environment of all particles. The snapshot is shown in the top row of Fig.\,\ref{fig:K2SO4}A (right one) with the particles in multiple colors signifying different kinds of environments in the system. The value of $r_c$ was $\sim$ 2.0 confirming the existence of 28 clusters (i.e. points due to the absence of any positional deviation) in the environment. Upon separating the particles with similar environment, the crystal class turned out to be Hexagonal with the requirement of further analyses to detect the unit cell and space group. The configuration of the Bravais lattice are presented in the top row of Fig.\,\ref{fig:K2SO4}B (left one). The value of $r_c$ was 6.0 in this case indicating the existence of eight cluster of points for the primitive Hexagonal Bravais lattice. As the original system had no statistical noise, the tolerances $\mathcal{X}_{d}$ and $\mathcal{X}_{a}$ were set to zero. The basis vectors appeared to be unique in this system; $\vec{a}$ = [5.947, 0, 0], $\vec{b}$ = [2.973, 5.15, 0] and $\vec{c}$ = [0, 0, 8.375] and corresponding lattice parameters were 5.947, 5.947, 8.375, $90^{\circ}$, $90^{\circ}$, $120^{\circ}$. The unit cell and coordinates of effective particles were shown in the top row of Fig.\,\ref{fig:K2SO4}B (right one) and the space group turned out to be $P6_{3}/mmc$ satisfying the initially considered unit cell used for the replication purpose. This example also shows the validity of this methodology suggesting the applicability in various crystal structures irrespective of the complexity.

\begin{figure*}[!h]
	\centering 
	\includegraphics[scale=0.23]{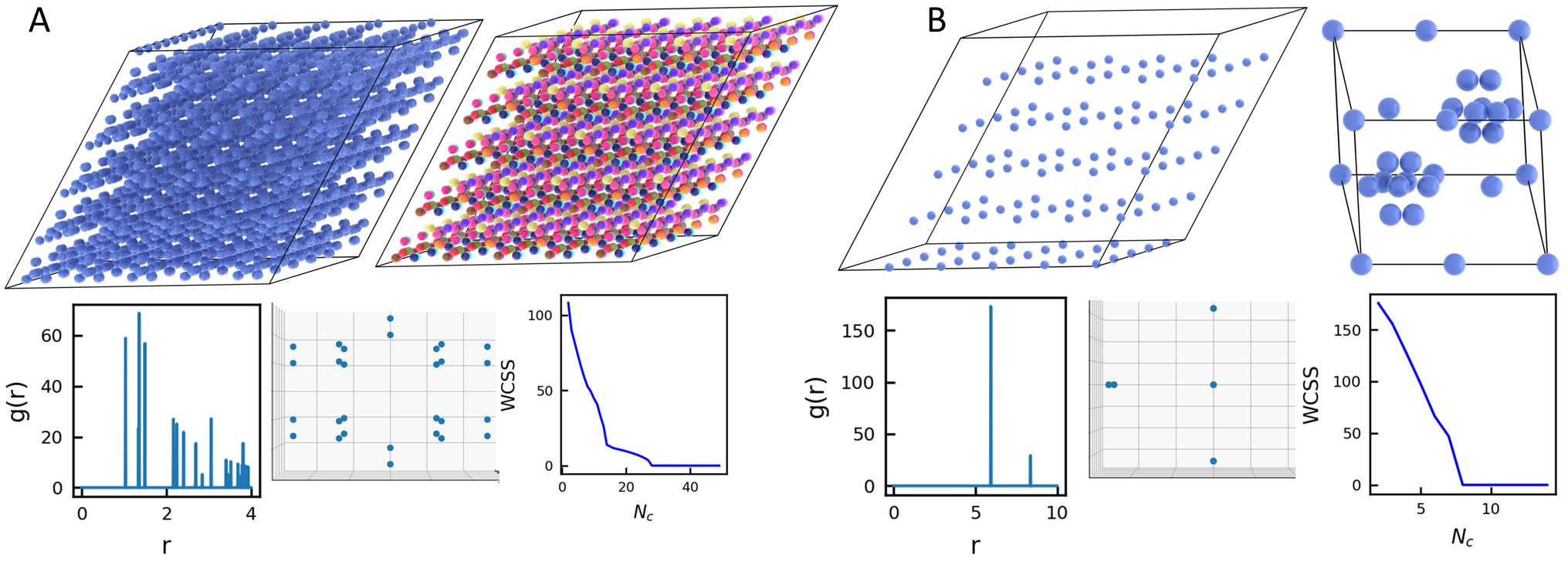}
	\caption{\textbf{The steps towards the detection of Hexagonal closed packed (HCP) structure of Potassium Sulfate (\ch{K2SO4}) are shown.} The coordinates of all atoms are shown in the top row of panel \textbf{(A)} (left figure). The cutoff distance for neighbors $r_c$ is chosen as 2.0 from RDF followed by the BOO calculation and estimation of clusters using \textsl{K-Means} clustering method as shown in the bottom row of panel \textbf{(A)}. The local environments of all particles are detected and the configuration is displayed in the top row of panel \textbf{(A)} (right one) with the particles in single color appear to have similar environment. The isolated particles in any single color are shown in the top row of panel \textbf{(B)} (left one). The similar analyses were applied on this system with $r_c$ $\sim$ 6.0 as presented in the bottom row of panel \textbf{(B)}. The detected unit cell with 22 effective coordinates is shown in the top row of panel \textbf{(B)} (right one) indicating the robustness of the algorithm to be effective in variety of crystal structures.}
	\label{fig:K2SO4}
\end{figure*} 

\newpage
\bibliography{unit_cell_detect_paper}

\end{document}